# Radiation damage assessment of SensL SiPMs

Lee Mitchell[1], Bernard Phlips[1], W. Neil Johnson[2], Mary Johnson-Rambert[1], Anika N. Kansky[3], Richard Woolf[1]

1. U.S. Naval Research Laboratory, 4555 Overlook Ave. S.W., Washington DC, 20375
2. Praxis Operations, 251 18th Street South, Suite 601, Arlington VA 22202
3. Naval Research Enterprise Internship Program (NREIP), 1818 N St NW, Suite 600, Washington, DC 20036

Email:lee.mitchell@nrl.navy.mil, PH# 202-404-7193

## Introduction

The work discussed here was performed to better quantify the current increase observed in Silicon photomultipliers (SiPMs) manufactured by SensL (now ON Semiconductor®) due to radiation damage [1]. SiPMs offer substantial size, weight and power savings and are quickly replacing traditional photomultiplier tubes (PMTs) as the preferred method for reading out scintillation detectors. SiPM technology is commonly used in commercially-available, terrestrial-based gamma-ray detection systems; however, spaced-based use of SiPMs for gamma-ray detection has only recently gained traction. Two recently launched, on-orbit gamma-ray instruments using SiPM technology are GRID [2], a Chinese Gamma Ray Burst (GRB) instrument utilizing SensL SiPMs and GAGG scintillator technology; and SIRI-1 [3], a U. S. Naval Research Laboratory mission to space qualify SiPMs and europium-doped strontium iodide ($SrI_2$:Eu). GRID launched on Oct 29, 2018 and SIRI-1 launched on Dec 3, 2018. SIRI-1 on STPSat5 finished its one year mission in a sun-synchronous orbit on Dec 3, 2019. The SIRI-1 instrument observed a 528 μA increase in the SiPM bias current over the one-year mission. A typical 2x2 J-series 60035 SiPM array operating at 28.9V, like the one on SIRI-1, has a dark current of roughly 10 μA. We have attributed this current increase to radiation damage.

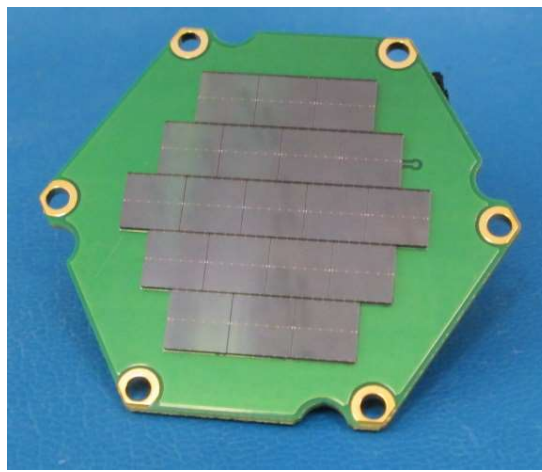

**Figure 1.** Readout board with 19 6 mm SensL SiPMs for a hexagonal $SrI_2$:Eu detector used on SIRI-2.

Future NRL missions using SiPM readouts are SIRI-2 [4], the follow on to SIRI-1, Glowbug [5] and the GAGG Radiation Instrument (GARI). Future NASA missions using SiPM technology include BurstCube [6], a GRB detection instrument and the CsI calorimeters on the gamma-ray observatory, AMEGO [7]. These missions cover a wide range of orbits and experience very different radiation fields. Both GARI and Glowbug are expected to launch in late 2021 and 2023, respectively, and will operate from the International Space Station (ISS) which is in a low earth orbit (LEO) at inclination of 51.6° and an altitude of 400 km. SIRI-2 is expected to launch in late 2020 into a geosynchronous orbit. Figure 1 shows a SiPM readout board for the hexagonal $SrI_2$:Eu detector to be used on SIRI-2. Depending on the voltage, a board with 19 SiPMs will draw ~50 µA of current at 28V. However, when damaged by radiation they can consume several mA of current. The power consumption is no longer negligible in the system's circuitry. Small PCB-mounted commercially available SiPM power supplies typically operate at a maximum of 4-10mA which will limit the number of SiPMs one can effectively bias from a single power supply. While there are other SiPM properties that are affected by radiation damage, the work here is focused solely on the issue of increased current.

General information on the design and operation of SensL SiPMs can be found elsewhere [1] [8]. Numerous irradiation studies with energetic particles show that radiation damage leads to increased bulk currents in SiPMs [9] [10] [11]. While much of the literature explores the performance of SiPMs for use in the harsh radiation environment of high-energy particle accelerators, the information is also applicable to space [12]. Future missions will involve large volume detectors with significantly more SiPMs needed to read out the scintillation light.

**Radiation Damage in Silicon**

Radiation damage in silicon detectors has been studied extensively for years and a more detailed summary is presented elsewhere [13] [14]. The types of damage can be broken down into two categories; bulk damage due to Non-Ionizing Energy Loss (NIEL) and surface damage due to the Ionizing Energy Loss (IEL) [15]. In general, low energy photons and electrons below (~300 keV) are limited to surface damage effects. Higher energy electrons, protons, neutrons and photons can generate significant damage in the bulk or volume of the material (bulk damage). With sufficient energy a particle can remove a Si atom from its lattice position. This is displacement damage and a sufficiently energetic particle can generate a number of displacements in the bulk of the material.

The defects generated by displacement damage are characterized by two types; point defects and cluster defects. The minimum energy to dislocate one Si atom and create a point defect is 25 eV. The silicon atom, the Primary Knock-on Atom (PKA), becomes an interstitial and leaves behind a vacancy and the two are commonly referred to as a Frenkel pair. When sufficient energy is transferred to the PKA, it can generate a cluster, a localized region of damage with many dislocations (displacements) and vacancies at the end of its track. In general these defects create energy states within the bandgap of the material. These traps and recombination centers lead to an increase in dark current. The bottom two rows of Table 1 show the energy threshold for creating these defects, denoted by $E_{min}$ point and $E_{min}$ cluster for a given particle. The first two rows show the maximum energy $K_{max}$ and the average energy $K_{avg}$ that a 1 MeV particle can transfer to a PKA. All values are in keV and are determined by the kinematics of the interaction [14].

| Radiation | e | p | n | Si+ |
|---|---|---|---|---|
| $K_{max}$ (keV) | 0.155 | 133.7 | 133.9 | 1000 |
| $K_{avg}$ (keV) | 0.046 | 0.210 | 50 | 0.265 |
| $E_{min}$ point (keV) | 260 | 0.190 | 0.190 | 0.025 |
| $E_{min}$ cluster (keV) | 4600 | 15 | 15 | 2 |

**Table 1.** Kinematic properties of 1 MeV particles in Si from Lutz [14]. All values are in keV.

The amount of displacement damage is proportional to the NIEL, and is referred to as the NIEL hypothesis. In general, one can compare the effects of different particle types and energy using the hardness factor K. Traditionally, the irradiation fluence, whether the particles are mono energetic or have a spectral distribution, is converted to a 1 MeV neutron equivalent to allow for easy comparison.

Figure 2 plots the NIEL and displacement damage cross-section, D, normalized to the 1 MeV neutron equivalent as a function of energy for protons, electrons and neutrons [16] [17] [18] [19]. Figure 2 can be used to determine the hardness factor, K, for a specific particle type and energy. For Si the conversion factor between NIEL and D is 100 MeV*mb=2.144 keVcm$^2$/g. To normalize the direct damage, the values in the secondary y-axis are divided by the 1 MeV neutron displacement damage cross section, $D_n(E=1MeV)=95 MeV*mb$. The hardness factor, K, of a mono energetic particle is defined by Equation 1, where the NIEL of the particle is given by $NIEL_p$, NIEL of a 1 MeV neutron is given by $NIEL_{En=1MeV}$, and the particle and 1 MeV neutron equivalent fluence are given by $\varphi_p$ and $\varphi_{En=1MeV}$

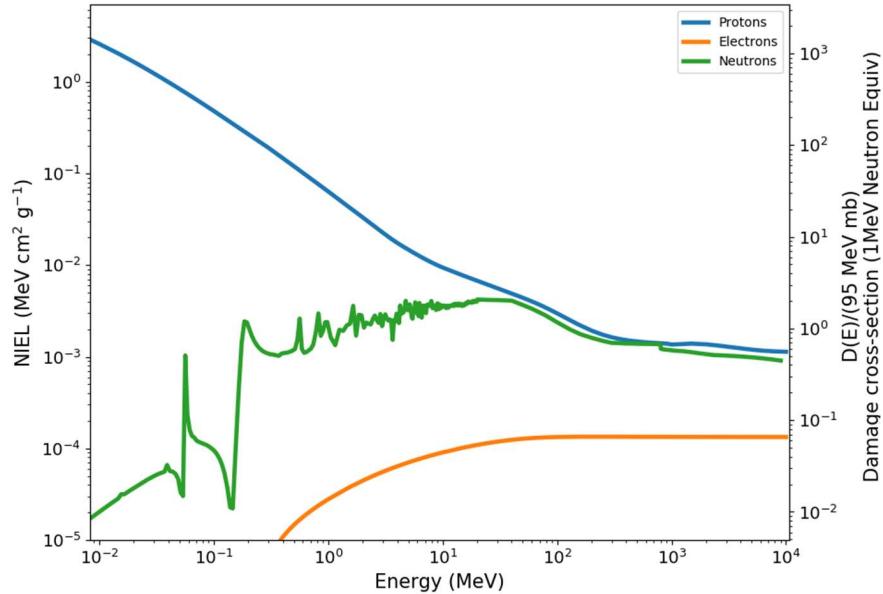

**Figure 2.** Plot of the NIEL and the displacement damage cross-section for various particles. Note the displacement damage has been normalized by the displacement damage of a 1 MeV neutron. The hardness factor, K, is given by the right axis [16] [17] [18] [19].

respectively.

$$K = \frac{NIEL_p}{NIEL_{n(E=1MeV)}} = \frac{D_p(E)}{D_n(E=1MeV)} = \frac{\varphi_{n(E=1MeV)}}{\varphi_p} \qquad \text{Eq. 1.}$$

Equation 2 can be used to determine the hardness factor for particles with a spectral energy distribution such as those found in the Van Allen radiation belts. Tabulated $D_p(E)$ values for many particle types may be found here [20] [21].

$$K = \frac{1}{D_n(E=1MeV)} * \frac{\int_0^{Emax} D_p(E)\varphi_p(E)dE}{\int_0^{Emax} \varphi_p(E)dE} \quad \text{Eq. 2.}$$

## Experimental Setup

### *General*

Table 2 list the number of SiPMs tested in an array of a given type and the physical properties of the SensL SiPMs tested. The break down voltage, $V_b$, was stated by the manufacturer to range between 24.3-24.7 V for the different SiPMs types. For our purposes we considered $V_b$ constant at 24.5 V at room temperature across all devices. Active area, cell size and fill factor of the various SiPMs are also listed in Table 2. Note: SiPM manufactures refer to the total area of the SiPM as the active area. This is different than the traditional use of active area in terms of radiation detection terminology. Only a fraction of the SiPM area is actually photosensitive. This fraction is the fill factor. In general, the smaller the cell size the smaller the fill factors as each cell requires its own isolation, quench resistor and signal tracks.

| Product # | Number of SiPMs Tested Electron (8MeV) | Number of SiPMs Tested Proton (64 MeV) | Active Area (mm²) | Cell Size (µm) | Fill Factor (%) |
|---|---|---|---|---|---|
| FC10010 | 15 | 15 | 1 | 10 | 28 |
| FC30035 | 10 | 10 | 9 | 35 | 64 |
| FC60035 | 8 | 8 | 36 | 35 | 64 |
| FJ60035 | 1 | 16 | 36 | 35 | 75 |
| FC30050 | 15 | 15 | 9 | 50 | 72 |
| FC30020 | 10 | 10 | 9 | 20 | 48 |
| FC10035 | 10 | 10 | 1 | 35 | 64 |
| FJ30020 | N/A | 8 | 9 | 20 | 62 |
| FJ30035 | N/A | 8 | 9 | 35 | 75 |
| RB10035 | N/A | 15 | 1 | 35 | 76 |

**Table 2.** List of SiPM products tested, along with number of SiPMs, active area, cell size and fill factor.

Proton and electron irradiations of the SiPMs were conducted at two different facilities. The 25 MeV S band LINAC at the Idaho Accelerator Center was used for 8 MeV electrons and the Crocker Nuclear Laboratory Cyclotron at UC Davis was used for 64 MeV protons. The choice of energy was largely determined by what could be achieved at each of the facilities given the desired fluence range.

*Electron Irradiation*

Figure 3 shows the target setup used during electron irradiation. For this test we irradiated the 7 types of SensL SiPMs listed in Table 2. The beam was 10 cm in diameter and considered uniform over this area. The charge/pulse was measured at $9x10^{-10}$ nC corresponding to $5.63x10^9$ electrons/pulse. Calibration runs were performed prior to the irradiation to ensure a inform dose. Dosimetry was used to determine that the total charge/pulse resulted in a fluence of $1.69x10^6$ electrons/cm$^2$/pulse on the target at a distance of 81 cm from the diffuser. A tungsten diffuser was used to ensure a uniform beam flux at this distance. The mono-energetic electron beam was set at 8 MeV which has a *dE/dx* of 1.916 MeV*cm$^2$/g in Si. Irradiation fluences ranged from $2.46x10^7$-$2.46x10^{11}$ electrons/cm$^2$. Table 3 lists the individual and cumulative fluences and dose.

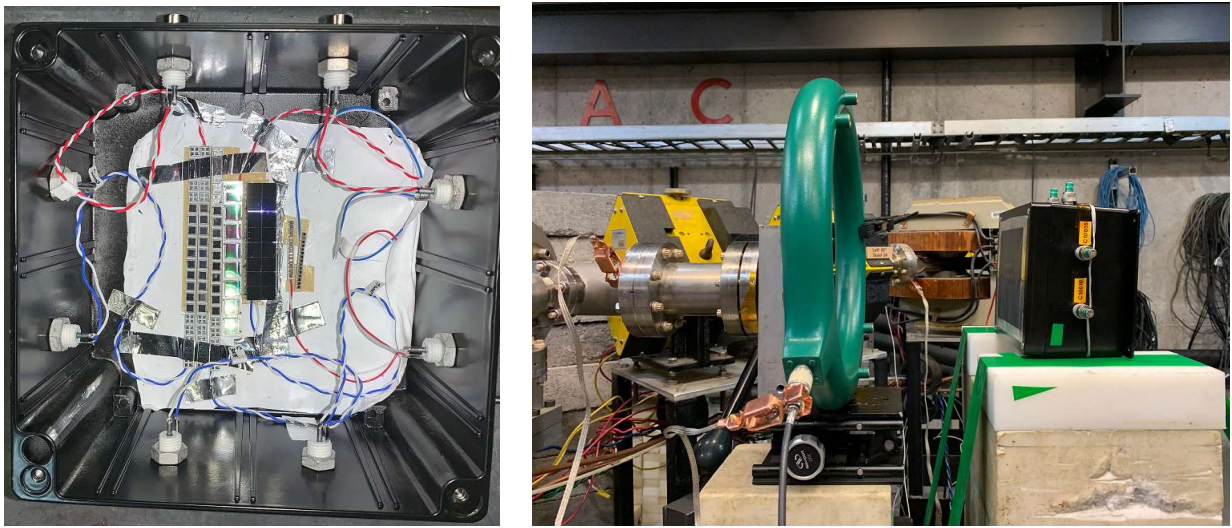

**Figure 3.** (left) shows the test box used to house the SiPM arrays and the experimental setup in front of the LINAC beam port at the Idaho Accelerator Center.

Current versus voltage measurements were made between each successive irradiation. A BeagleBone Black running a Python script was used to scan an AiT HV80 SiPM power supply. The accuracy of the AiT HV80 to measure low currents is reflected in the error bars for low current densities. We compensated for this by adding more SiPMs and averaging over an array. In most cases, we irradiated an array of SiPM and measured the total current dividing by the number of SiPMs. An error in wiring that was not detected prior to the irradiation measurements meant we only measured one FJ60035 SiPM during the experiment and is noted in Table 2. The AiT HV80 is the same power supply used on SIRI-1 and is capable of 4 mA total current. Current densities were measured between 25 and ~30V, the operational range of the SensL SiPMs. In hindsight we should have used a more versatile power supply than the AiT HV80, which had limited resolution and range. This was the primary reason we switched to the Keithley Power Supply (K237) used in the proton irradiation studies. All measurements were performed at room temperature.

|  | 8-MeV Electron |  |  | 64-MeV Protons |  |
|---|---|---|---|---|---|
| Incremental Fluence e/cm² | Cumulative Fluence e/cm² | Cumulative Dose rad(Si) | Incremental Fluence p/cm² | Cumulative Fluence p/cm² | Cumulative Dose rad(Si) |
| 1.55E+07 | 1.55E+07 | 4.77E-01 | 1.28E+06 | 1.28E+06 | 1.71E-01 |
| 6.22E+07 | 7.77E+07 | 2.39E+00 | 1.30E+06 | 2.58E+06 | 3.44E-01 |
| 1.55E+08 | 2.33E+08 | 7.16E+00 | 3.74E+06 | 6.32E+06 | 8.44E-01 |
| 3.10E+08 | 5.43E+08 | 1.67E+01 | 5.04E+06 | 1.14E+07 | 1.52E+00 |
| 1.55E+09 | 2.10E+09 | 6.44E+01 | 1.11E+07 | 2.24E+07 | 3.00E+00 |
| 3.10E+09 | 5.20E+09 | 1.60E+02 | 3.98E+07 | 6.22E+07 | 8.31E+00 |
| 6.22E+09 | 1.14E+10 | 3.51E+02 | 6.23E+07 | 1.25E+08 | 1.66E+01 |
| 1.55E+10 | 2.70E+10 | 8.28E+02 | 1.25E+08 | 2.49E+08 | 3.33E+01 |
| 3.10E+10 | 5.80E+10 | 1.78E+03 | 3.98E+08 | 6.47E+08 | 8.64E+01 |
| 6.22E+10 | 1.20E+11 | 3.69E+03 | 5.97E+08 | 1.24E+09 | 1.66E+02 |
| 1.55E+11 | 2.76E+11 | 8.46E+03 | 1.25E+09 | 2.49E+09 | 3.33E+02 |
|  |  |  | 3.98E+09 | 6.47E+09 | 8.64E+02 |
|  |  |  | 5.97E+09 | 1.24E+10 | 1.66E+03 |
|  |  |  | 1.24E+10 | 2.49E+10 | 3.32E+03 |
|  |  |  | 3.99E+10 | 6.47E+10 | 8.64E+03 |

**Table 3.** Fluence and dose for both electron and proton irradiations.

*Proton Irradiation*

For proton irradiation; the SiPMs were enclosed in a light tight box with a 2 mil Tedlar polyvinyl fluoride (PVF) window as shown in Figure 4 (left). The actual beam port has a thin Kapton window to maintain vacuum and with the collimator generates a 7.6-cm-diameter beam. Figure 4 (right) shows the test box directly in front of the beam port. All devices were grounded to the sample stage during the irradiation process. The measurements were collected in the North Cave of the UC Davis facility, Figure 5 (left). The SiPM arrays were arranged to fit within a 5 cm diameter allowing for easy alignment and reducing the risk of encountering the beam edge. The number of SiPMs for each array may be found in Table 2. The beam profile is shown in Figure 5 (right). Alignment lasers provided by the facility were used to insure the setup was centered with respect to the beam. The outer edge of the beam flux on our SiPM targets was measured to be 97.5% of the center. For the purposes of this paper we consider the beam uniform and all SiPM saw the same fluence. The proton beam energy was 64 MeV which has a *dE/dx* of 8.334 MeV*cm²/g in Si.

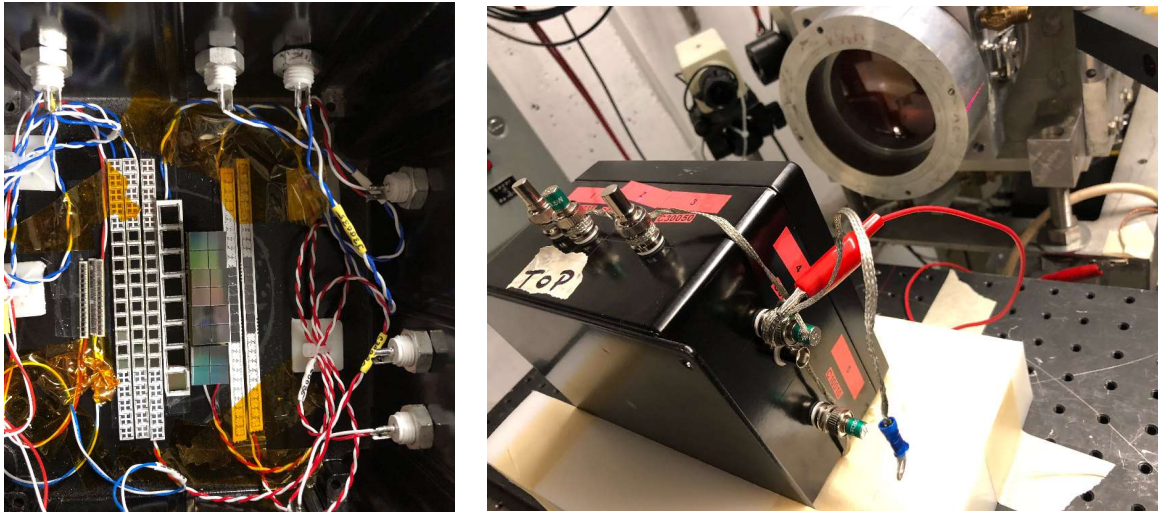

**Figure 4. (left)** Photo of open test box showing the arrangement of the various SiPM arrays. **(right)** Photo experimental testbox showing the test box directly in front of the aperture.

Table 3 lists the incremental and cumulative fluence as well as the total dose. Current-voltage (IV) measurements were made between each successive irradiation using a Keithly 237 High Voltage Source Measurement Unit. Voltage measurements were made over the range 24-30V in increments of 0.5V. Since our beam time was limited to approximately ½ a day, we wanted to facilitate rapid turnaround between irradiations. Measurements were made using the 0-10mA current scale where the instrument is limited to 700 nA of resolution. This allowed us to cover a larger dynamic range in current measurements, but resulted in larger error bars at low current densities for some devices. All measurements were performed at room temperature.

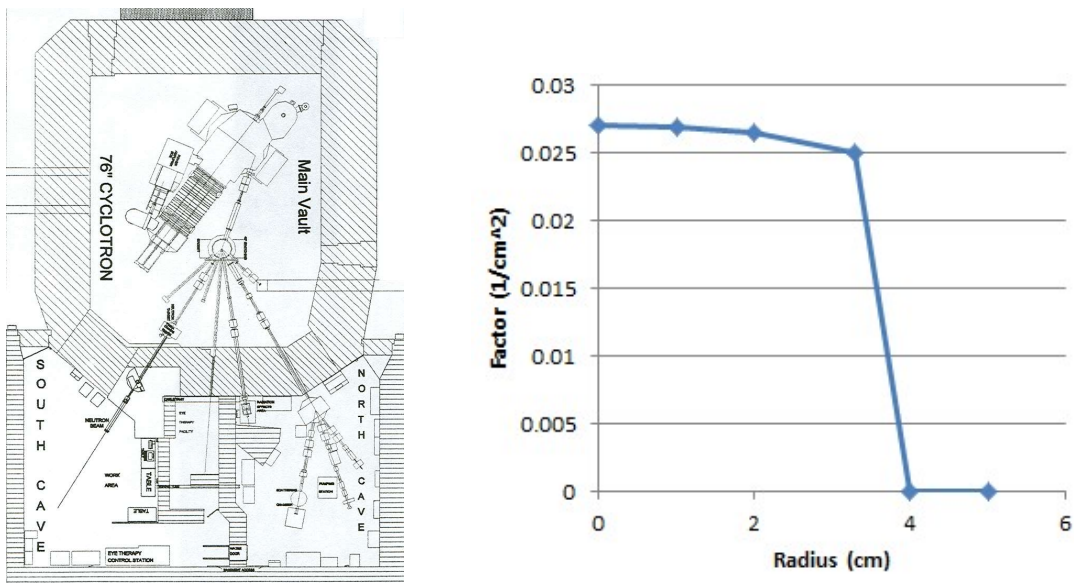

**Figure 5. (left)** Map of UC Davis Cyclotron Facility. **(right)** Radial intensity of proton beam at target.

## Discussion of Results

All data discussed in the results section are plotted a function of current density as defined by Equation 3, where $J$ is the current density in µA/cm², $I$ is the measured current in µA, $N$ is the number of SiPMs in the measured strand, A is the active area of one SiPM in cm² and $FF$ is the fill factor as defined by the manufacturer. Normalizing the measured values resulted in the grouping of devices by cell size, and in general the larger the cell size the more susceptible the SiPM. This is discussed in more detail in the following sections.

$$J = \frac{I}{A*N*FF} \quad \text{Eq. 3.}$$

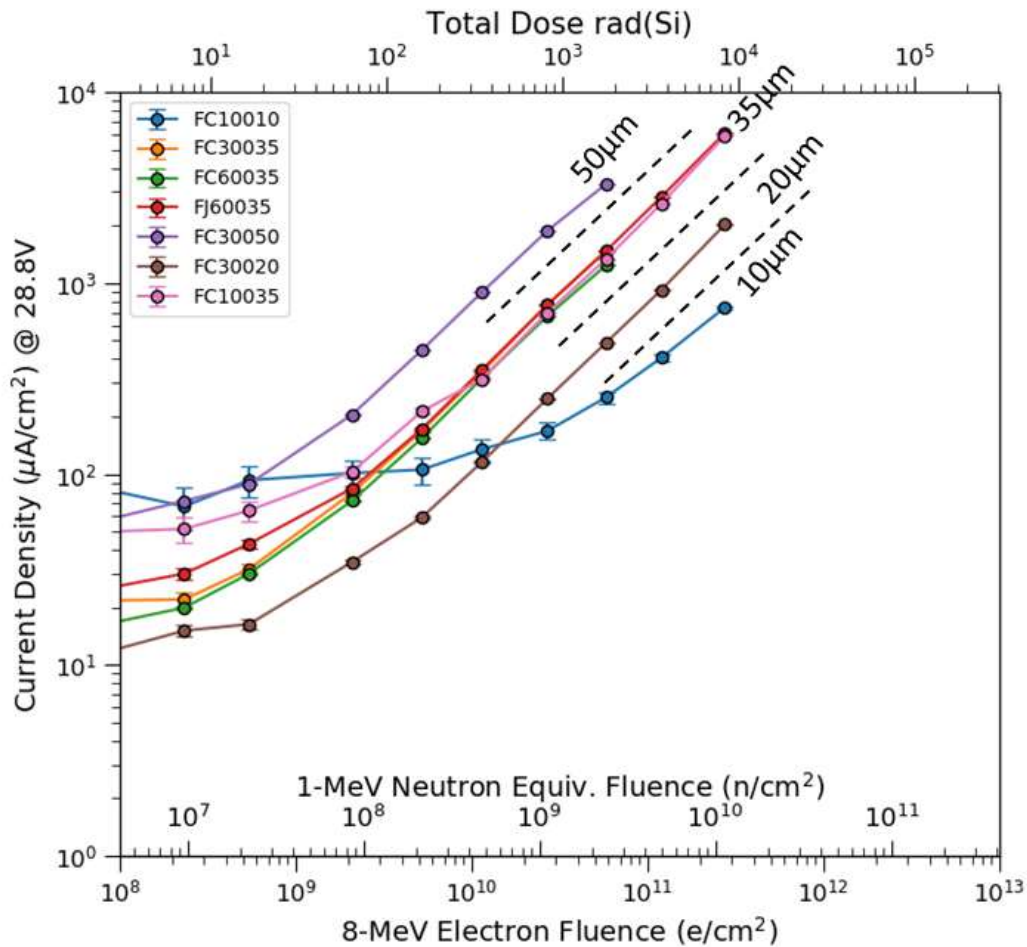

**Figure 6.** Plot of current density as a function of dose/damage of for all SiPMs biased at 28.8V. SiPMs with the same cell size showed comparable current densities. In general the larger cell size SiPMs showed more susceptibility to radiation damage from a bulk current standpoint. Data collected from 8 MeV electron beam.

*Electron Irradiation*

Traditional I-V plots for each SiPM and fluence may be found in Appendix A. Figure 6 shows a plot of the current density for all tested SiPMs as a function of total dose, the 1MeV equivalent neutron fluence and the electron fluence. For conversion to 1-MeV neutron equivalent we used a NIEL value of 8.4161e-05 MeV*cm$^2$/g. Here the current density is defined by Eq. 3. All SiPMs were biased at 28.8V for this plot. Appendix B provides these same plots at other measured voltages. Figure 6 shows that SiPMs with the same cell size have comparable current densities as a function of fluence and is independent of series for the two types measured here. SiPMs with larger cell size showed a larger current density for a given fluence. For comparison we take the current density measured at $10^9$ n/cm$^2$ (1MeV Eq.) for 35μm SiPMs and find the value to be ~800 μA/cm$^2$. This compares well with Figure 7 which shows the equivalent measurement of current density after proton irradiation. Given the same neutron equivalent fluence a current density of ~840 μA/cm$^2$ was measured on the 35 μm SiPMs and was expected based on the NIEL hypothesis.

*Proton Irradiation*

Traditional I-V plots for each SiPM and fluence may be found in Appendix A. Figure 7 shows a plot of the current density for all tested SiPMs as a function of total ionizing dose, the 1 MeV equivalent neutron fluence and the actual fluence. For conversion to 1-MeV neutron equivalent, we used a NIEL value of 3.8892e-03 MeV*cm$^2$/g. All SiPMs were biased at 28.5V for this plot. Appendix B provides these same plots for other measured voltages. The plot shows that SiPMs with the same cell size have comparable current densities as a function of fluence, with the exception of FC30035 and RB10035. Our lack of information on the specifics of the manufacturing process (layer thickness) and electrical properties (i.e., depletion width) make it difficult to know exactly which device parameter makes certain devices more sensitive to radiation damage. Nevertheless, the SensL FC and FJ series SiPMs showed consistent current densities for a given cell size, with the larger cell sizes showing more susceptibility to radiation damage. FC30035 was the exception and we expected it to follow the trend of the other 35 μm SiPMs. However, at higher voltages the damage curve begins to move towards the 35 μm band (Appendix D, Figure D11). After the irradiation, we verified the readout board was populated with correct SiPMs, and were unable to identify the cause of this effect. Secondly, the same model of SiPM was irradiated with electrons (shown in the previous section) and compared well with other 35 μm SiPMs. We understand the FC and FJ series use a P$^+$-on-N diode structure which makes them more efficient in the shorter wavelengths and the majority of the depletion region extends into n-type silicon [22]. The fact we observed comparable results across series is not surprising. On the other hand RB10035 (a red sensitive SiPM) is a N$^+$-on-P diode structure with a poly coating on top making them more efficient to longer wavelengths. The depletion region extends into the p-type Si. In general it showed a lower susceptibility than other SiPMs [23]. The fabrication and device characteristics of the RB series are expected to be significantly different than the FJ and FC series.

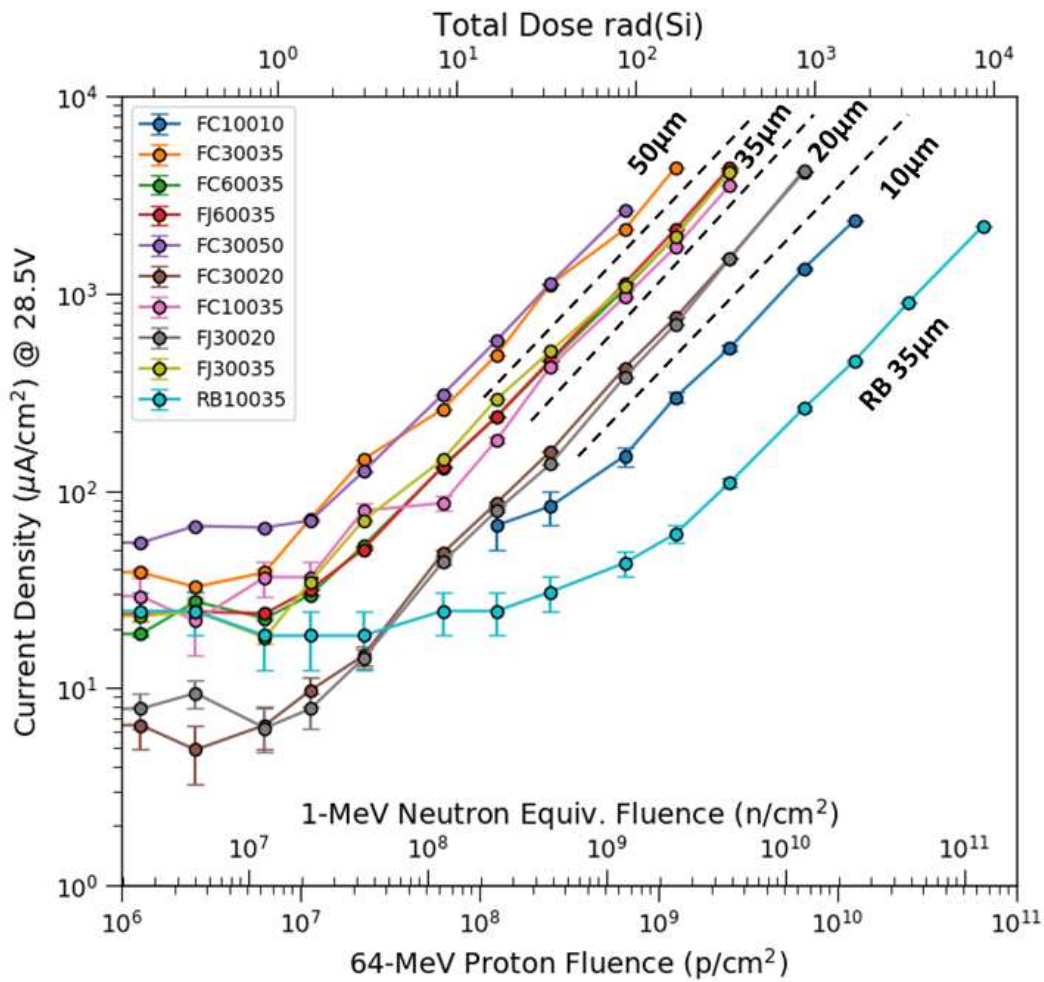

**Figure 7.** Plot of current density as a function of dose/damage of for all SiPMs biased at 28.5V. SiPMs with the same cell size showed comparable current densities, with the exception of FC30035 and RB10035. In general the larger cell size SiPMs showed more susceptibility to radiation damage from a bulk current standpoint. Data was collected with 64 MeV proton beam.

*Annealing Effects*

SiPMs are expected to experience some self-annealing over the life of the space mission. As other authors have observed, we see a sharp reduction in SiPM current during the first 10 days after irradiation [24]. Qiang shows a similar plot for an older model SensL SiPMs and fits the data to a single curve showing a time constant of ~10 days [25]. Annealing is a complicated process between different defects and is only partially understood. Lutz recommends parametrization of the annealing process in the form of Equation 4 for bulk current, where $t$ is time and $\tau$ is the time constant of the decreasing current [14]. Other authors have had success fitting the annealing data for SensL SiPMs over shorter periods of time using a single exponential decay and offset parameter. In this case they also found the time constant to have an inverse relationship with dose. Values ranged between 6-10 days for fluences between 15.5 and 0.150 krads [26], respectively.

$$\frac{\Delta I(t)}{\Delta I(0)} = A_0 + \sum_{i=1}^{n} A_i e^{\frac{-t}{\tau_i}} \quad \text{Eq. 4.}$$

In most cases, or at least it has been our experience that most instruments are kept at or near room temperature while in orbit. SIRI-1 maintained a temperature range between 6-9°C for most of its mission life. All samples were kept at room temperature (~22°C) since they were returned to NRL and the current was periodically measured over a 90 day time period. Plotted in Figure 8, the annealing curves for the various SiPMs all show a similar profile and appear to be only offset by their starting current. In the first ~10 days we see the current of the FJ60035 drop from 10,600 µA/cm² to 5,440 µA/cm², 49% of its original value. This portion of the curve is comparable to the results shown in [25]. Since that time the annealing process has slowed considerably. After an additional 70 days of room temperature annealing the sample had a current of 3,790 µA/cm², only a 30% reduction. Unlike the irradiation test, the on-orbit radiation dose will occur slowly over time and the radiation damage and annealing will occur concurrently.

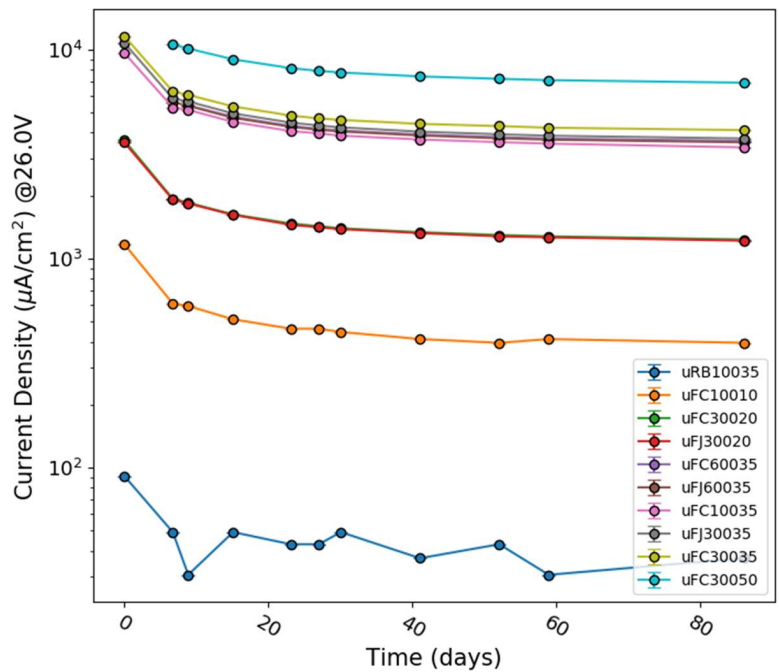

**Figure 8.** Plot of current density as a function of annealing time for all proton irradiated SiPMs. Samples were kept at room temperature between measurements. The one exception to this was during shipping back to NRL.

Using the LMFIT fitting module in Python3, we attemped to fit the curves with a single exponitial decay and offset (Eq. 4 with *n*=1). Attempts to fit all points with a single exponential resulted in poor fits. Removing the first point (the last measurement immediately after irradiation) resolved this issue. Figure 9 shows a plot of the room temperature annealing data and the fit for FJ30035. This would be the case if different times of the annealing data were dominated by different time constants. One reason for this is the fact our instrument and electronics were shipped back to NRL appoximately 7 days after the intial iradiation. The instrumentation traveled back to NRL the first week in August 2019 from the Sacramento, California area. At the tmie daily temperatures peaked in the high 90s °F. Unfortunately, the temperature during this time was not recorded and the SiPMs could have experienced elevated temperatures without our knowlegde. Others have shown higher temperatures lead to faster recovery times (shorter time constants) [25]. Another possibility is that the annealing process is more complicated and a single decay exponential is not enough to define the process [14]. Fit parameters for all devices are shown in Table 4 and the time constant determined by fitting the points starting 7 days after the last irradiation was fairly consistance between SiPMs.

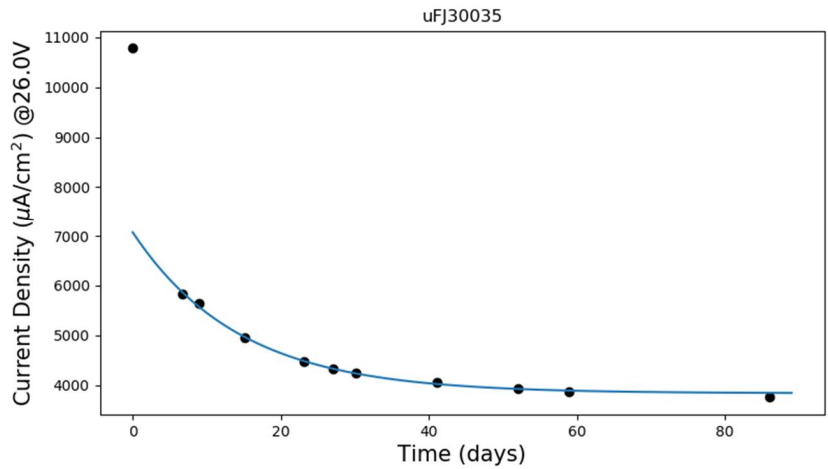

**Figure 9.** Plot shows a fit (blue line) of the annealing data for the SensL FJ30035 where current density is defined by Eq. 3. First point ($t=0$) is excluded from fit.

| SiPM Model | Amplitude (µA/cm²) | Time Constant (Days) | Offset(µA/cm²) |
|---|---|---|---|
| FC10010 | 343.7 +/- 20.7 | 14.7 +/- 1.92 | 395.85 +/- 9.03 |
| FC30020 | 1081.6 +/- 17.7 | 13.89 +/- 0.452 | 1269.75 +/- 6.65 |
| FJ30020 | 1075.6 +/- 19.8 | 14.02 +/- 0.52 | 1253.56 +/- 7.62 |
| FC60035 | 2725.9 +/- 178 | 16.25 +/- 1.11 | 3668.48 +/- 23.3 |
| FJ60035 | 3143.8 +/- 131 | 14.52 +/- 0.848 | 3645.18 +/- 28.6 |
| FC10035 | 2875.5 +/- 85.3 | 14.42 +/- 0.894 | 3518.65 +/- 35.1 |
| FJ30035 | 3237.6 +/- 73.4 | 14.38 +/- 0.679 | 3834.75 +/- 30 |
| FC30035 | 3458.8 +/- 73.5 | 13.69 +/- 0.569 | 4208.95 +/- 26.7 |

**Table 4.** Fit parameters for room temperature annealing data on all SiPM models for currents measured at 26V. (Only measurements 7 days after the irradiation were included in the fit.).

## On-Orbit Considerations

One can use trapped particle models AE9 and AP9 to generate an annual fluence or flux for a given orbit and particle type. Monte Carlo codes along with the mass model of the instrument and spacecraft can then be used simulate the fluence experienced by the detector's SiPMs over a given time frame. One can then use Equation 2 to generate a hardness factor for the energy distributed fluence. A comparison of estimated dose and the plots provided in the Appendix C&D can be used to predict the expected current increase on future missions. As an example we provide a comparison between the increases in measured current on SIRI-1 to a back of the envelope estimate we generate from the results of this paper. We opt to use the online application SPENVIS [27] for quick and easy estimates of the on-orbit radiation environment. First we provide SPENVIS with information about the orbit. In the case of SIRI-1, it was just a circular 600 km sun synchronous orbit. We use the AE9 and AP9 trapped particle models accessible through SPENVIS to generate an average flux based on this orbit and mission dates [28]. We neglect the smaller contributions to dose from cosmic-rays and other source. SPENVIS can then uses these results in SHEILDOSE2 and generate and estimate of annual dose as a function of shield thickness, Figure 10 [29]. For SIRI-1, the SiPMs have a 40 mm thick $SrI_2$ crystal on one side and the spacecraft below that provides substantial shielding. While the plot only goes to 20 mm (of Al shielding), we note the primary contributor to dose at these thicknesses are the protons and the curve asymptotically approaches 90-100 rads(Si) with increasing shield thickness.

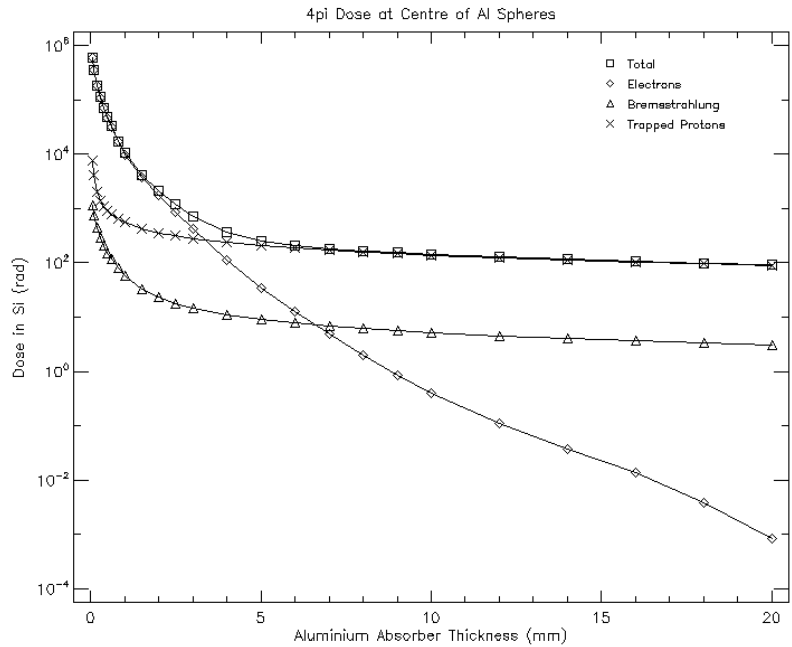

**Figure 10.** Plot of on-orbit annual dose as a function of Al shielding generated using SPENVIS and SHIELDOSE2 for the SIRI-1 instrument.

Here we assume the on-orbit energy distribution of protons will generate comparable damage to the 64 MeV protons used during the irradiation test. From Figure 2 we note there is only a small change in the displacement damage cross section for protons higher than 64 MeV making this assumption reasonable. Ideally, one would generate the 1 MeV neutron equivalent for their energy distribution of particles. From Figure 5 we get ~1000 µA/cm$^2$ for the FJ60035 at 100 rads(Si). If the current density is multiplied by the fill factor and active area, we get 270 µA per SiPM. This comes to a total of 1,080 µA for four SiPMs. The SiPMs will experience some annealing during this time. We use 60% as a rough estimate in the reduction in current due to room temperature annealing during the life of the mission and we expect 432 µA of current to be generated annually in our 2x2 array. We measured a value was 528 µA at the end of the SIRI mission. The predicted value was within 20% of the measured value and is reasonable given the number of approximations made.

## Conclusion

We have irradiated a wide range of SensL SiPMs with both energetic protons and electrons to characterize the increase in dark current as a function of fluence and over voltage. In both cases, we began to observe a current increase at a 1-MeV-neutron-equivalent fluence of ~2x10$^7$ n/cm$^2$, which roughly corresponds to 10$^7$ 8-MeV-electrons/cm$^2$ and 10$^9$ 64-MeV-protons/cm$^2$. When normalized with respect to active area and fill factor, results showed SiPMs with the same cells size performed similarly. In general it was observed, SiPMs with larger cell size showed higher bulk currents at the same fluence with the notable exception of the red-sensitive variant. The relationship between bulk current and fluence appears to be linear and our results are in agreement with the NIEL hypothesis. Results measured here also compared well with on-orbit observation from SIRI-1 and can be used to estimate the performance of future systems. Future work will measure the performance of SiPMs from other manufacturers and seek to compare results across brands.

**Acknowledgements**: This work was sponsored by the Chief of Naval Research. The authors would also like to thank Jon Stoner and the Idaho Accelerator Center staff and Carlos Castaneda and the UC Davis Cyclotron staff.

**Appendix A. I-V Curves for 8 MeV electron irradiated samples.**

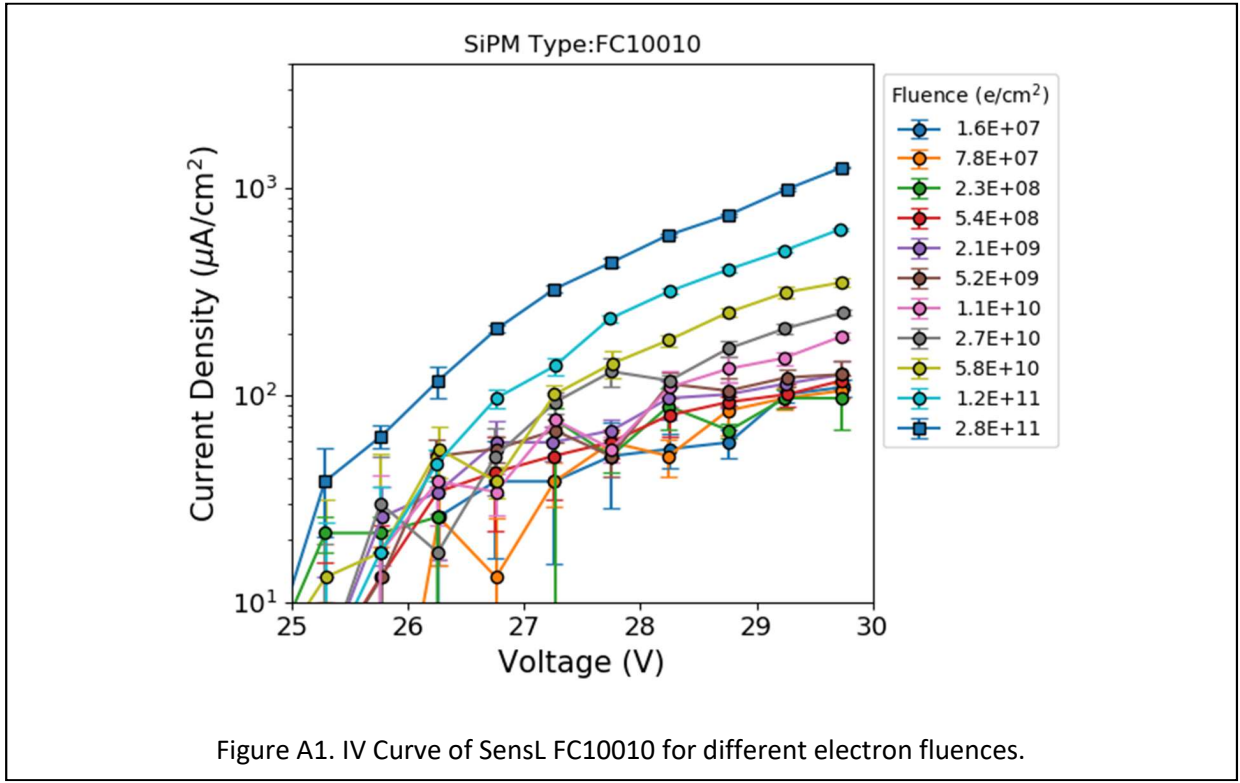

Figure A1. IV Curve of SensL FC10010 for different electron fluences.

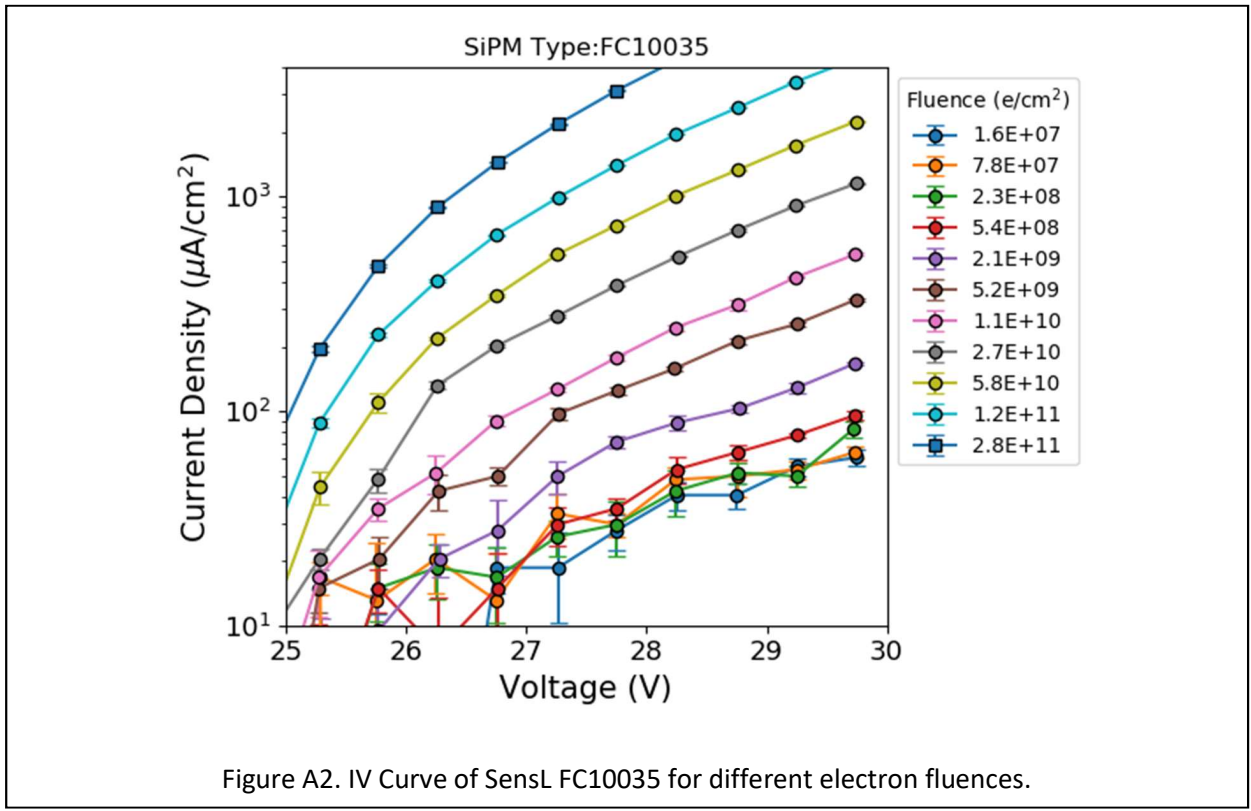

Figure A2. IV Curve of SensL FC10035 for different electron fluences.

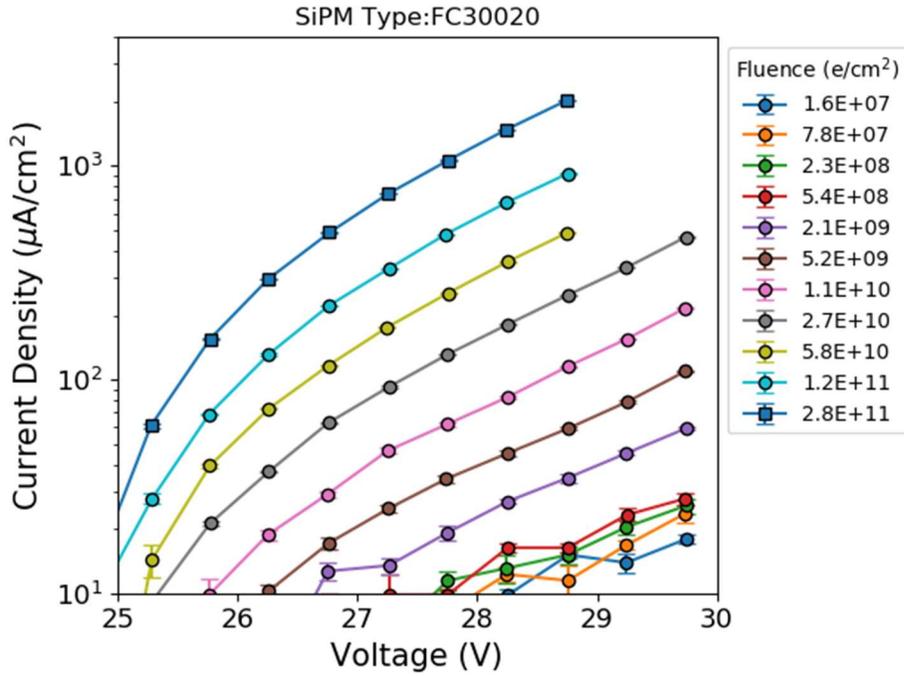

Figure A3. IV Curve of SensL FC30020 for different electron fluences.

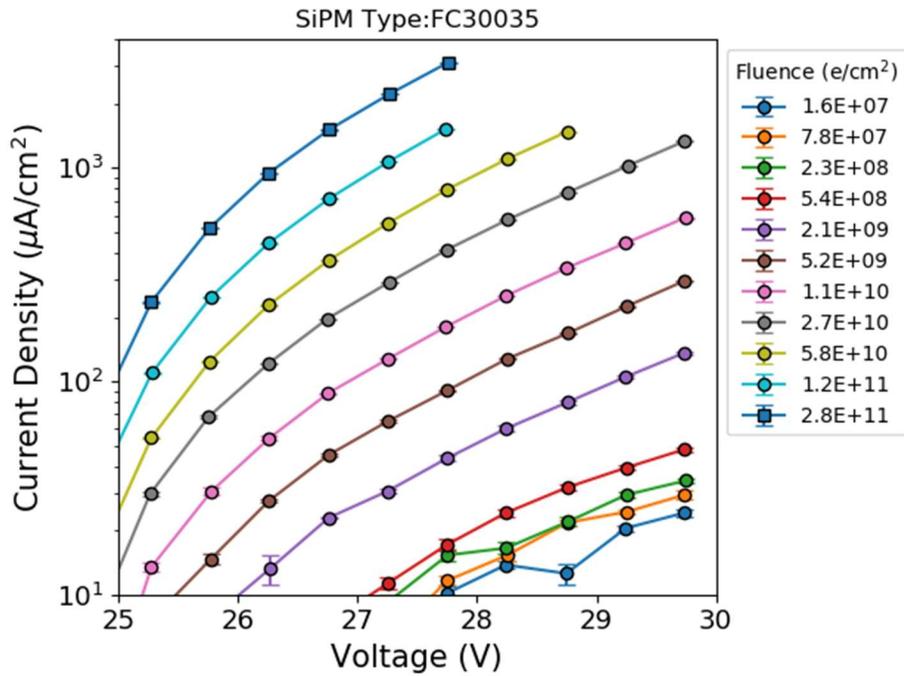

Figure A4. IV Curve of SensL FC30035 for different electron fluences.

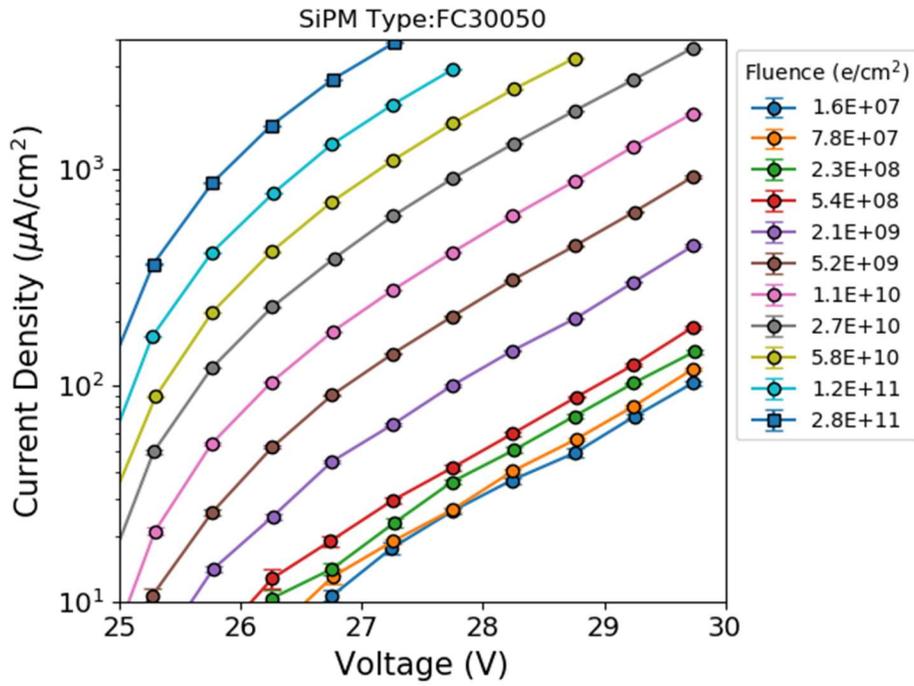

Figure A5. IV Curve of SensL FC30050 for different electron fluences.

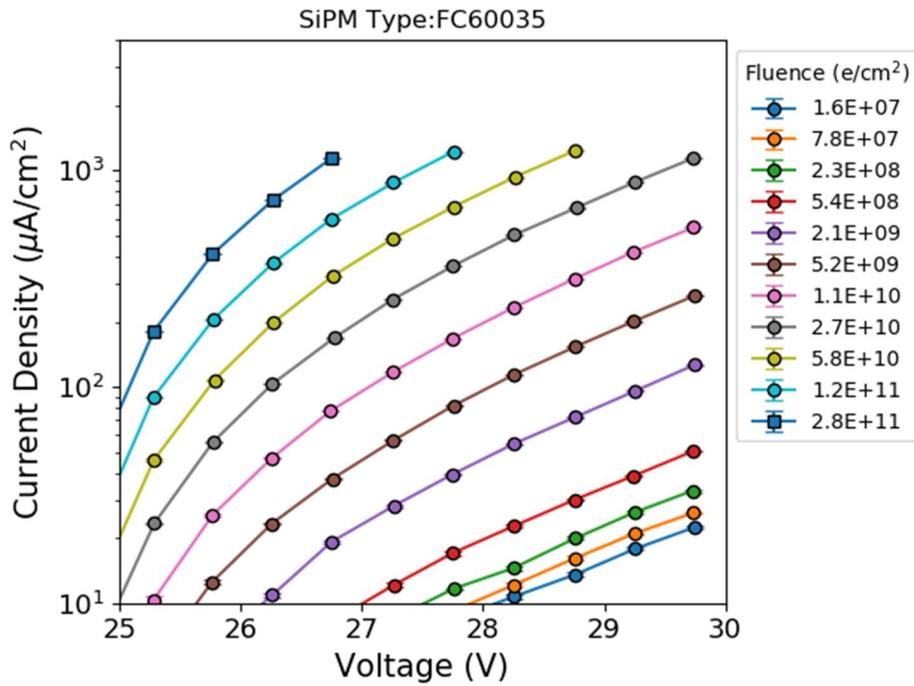

Figure A6. IV Curve of SensL FC60035 for different electron fluences.

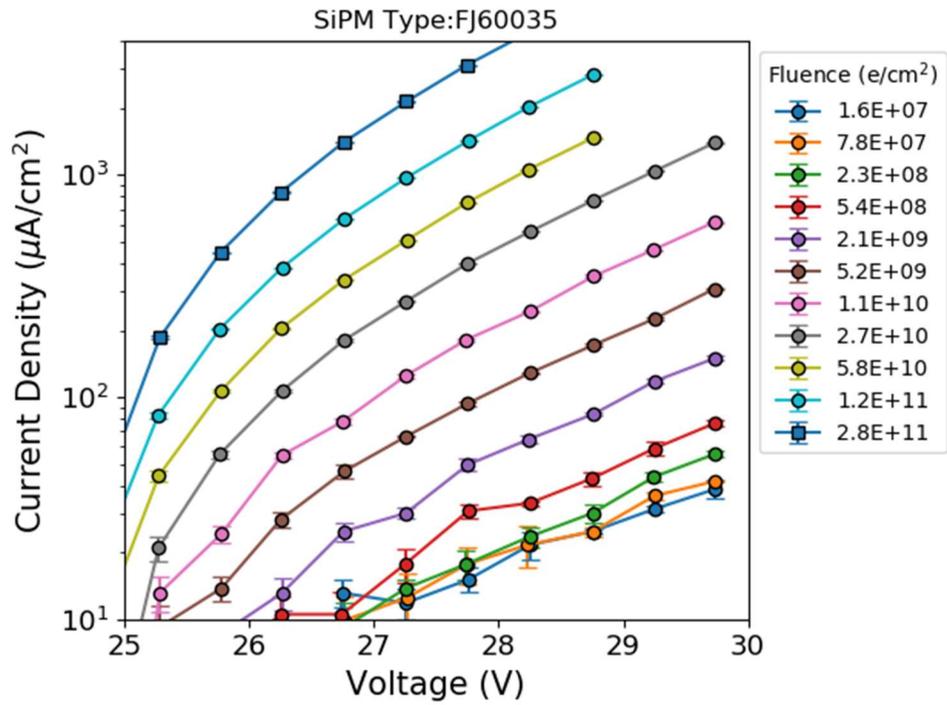

Figure A7. IV Curve of SensL FJ60035 for different electron fluences.

**Appendix B. Current density as a function of dose for 8 MeV electron irradiated samples.**

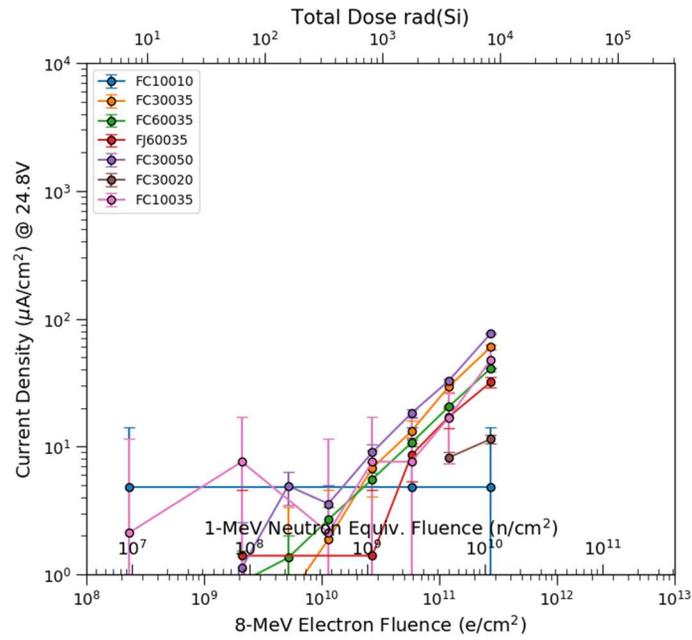

Figure B1. Current Density as a function of dose\fluence at 24.8V bias for alls SiPMs.

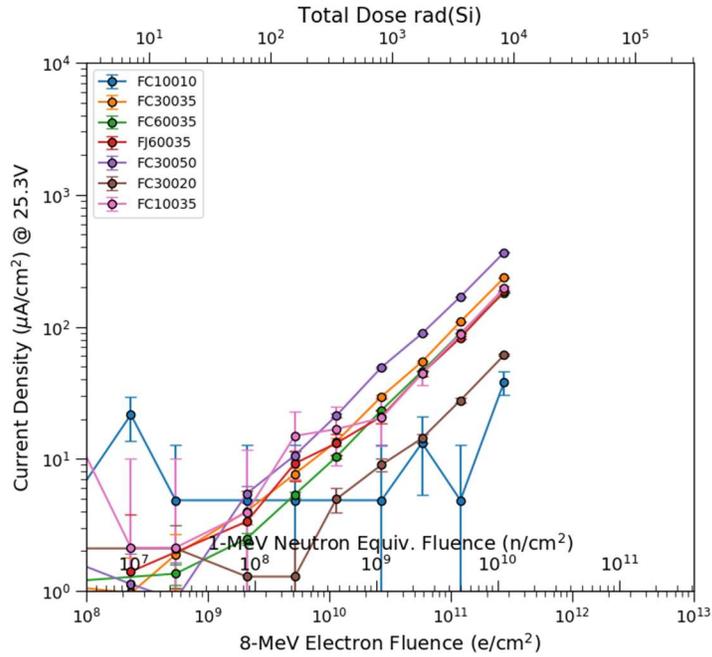

Figure B2. Current Density as a function of dose\fluence at 25.3V bias for alls SiPMs.

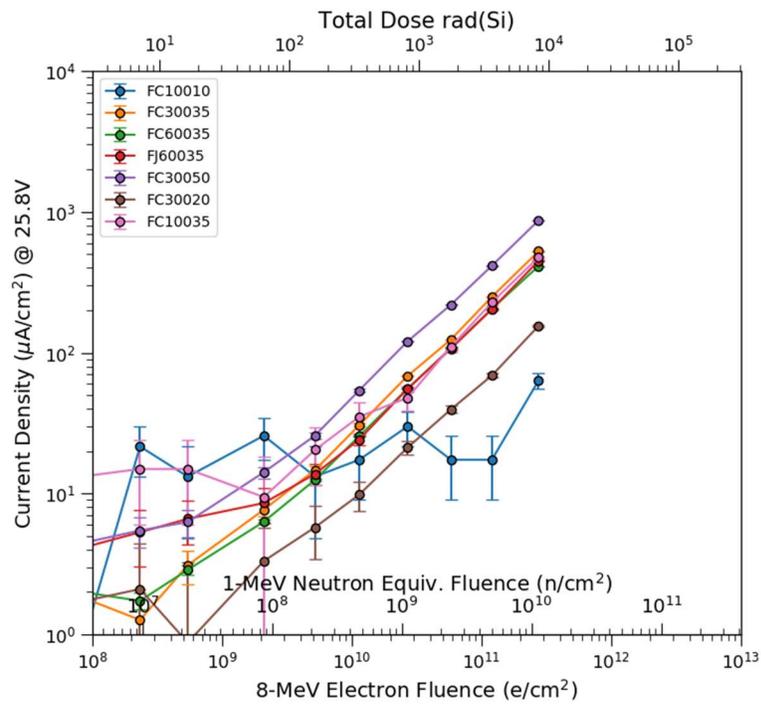

Figure B3. Current Density as a function of dose\fluence at 25.8V bias for alls SiPMs.

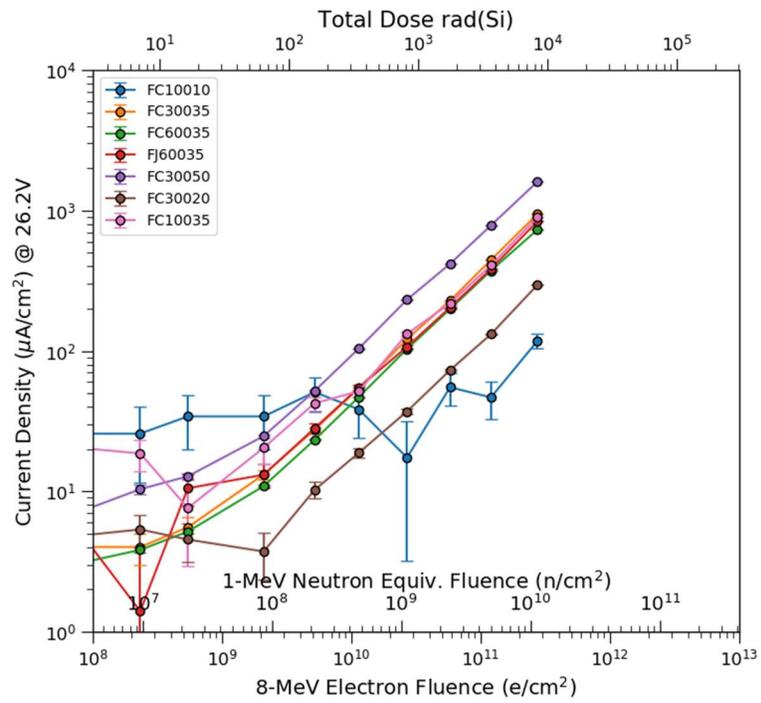

Figure B4. Current Density as a function of dose\fluence at 26.2V bias for alls SiPMs.

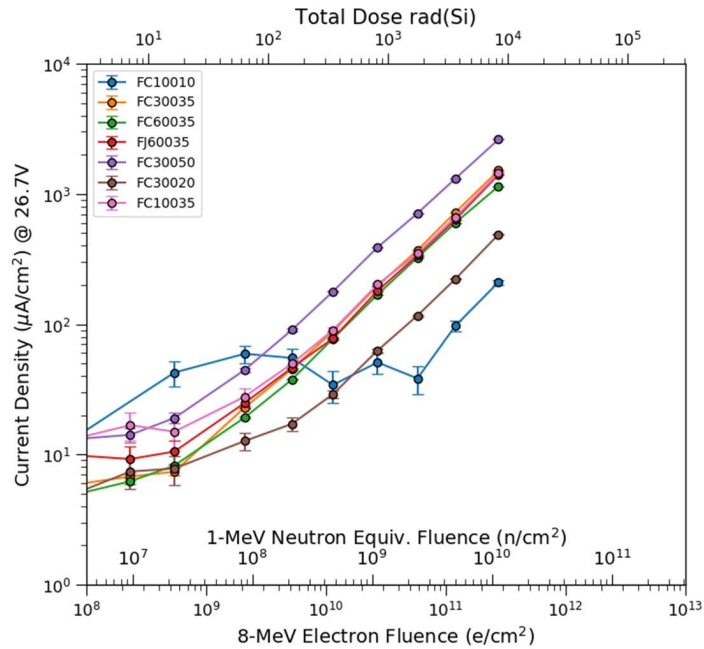

Figure B5. Current Density as a function of dose\fluence at 26.7V bias for alls SiPMs.

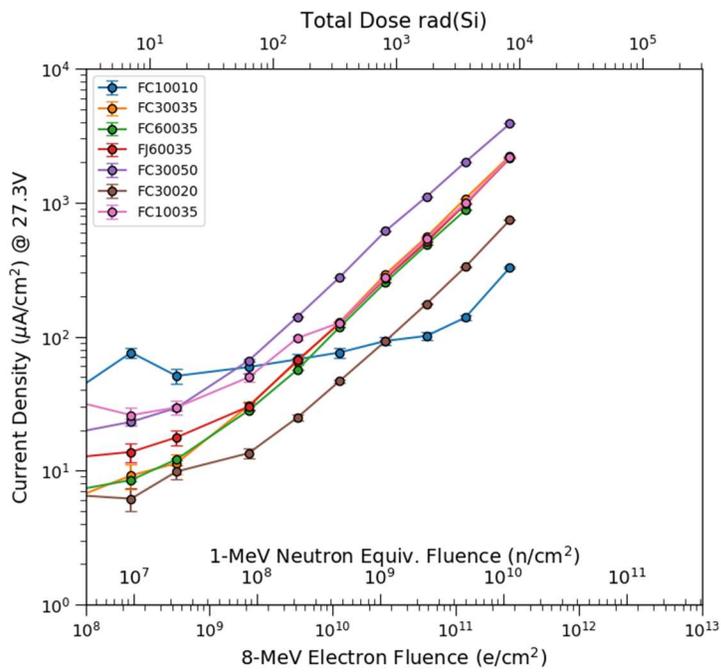

Figure B6. Current Density as a function of dose\fluence at 27.3V bias for alls SiPMs.

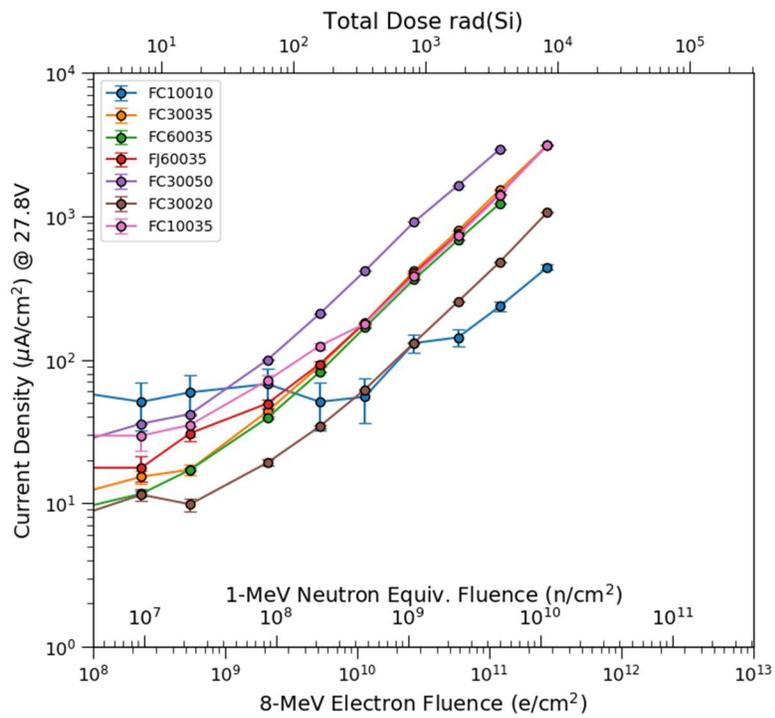

Figure B7. Current Density as a function of dose\fluence at 27.8V bias for alls SiPMs.

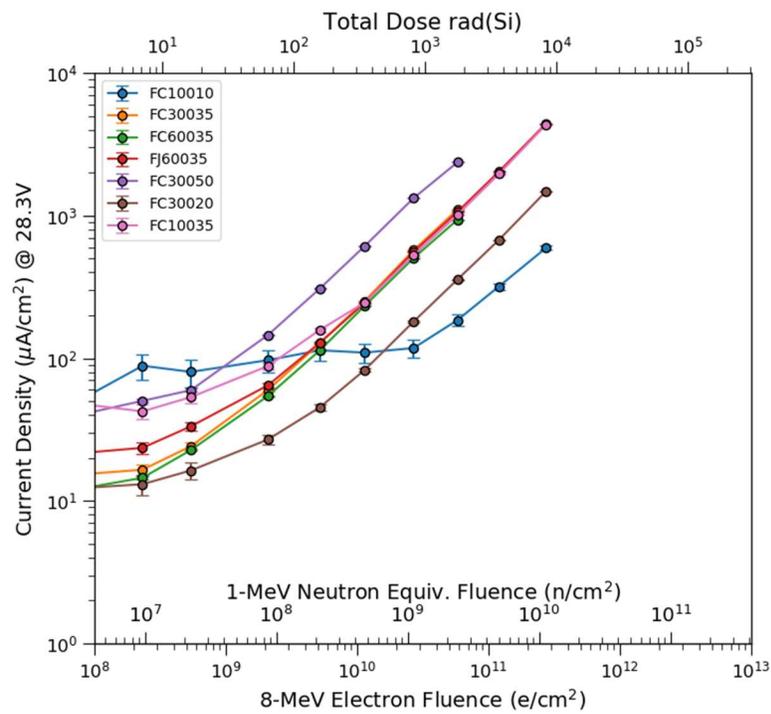

Figure B8. Current Density as a function of dose\fluence at 28.3V bias for alls SiPMs.

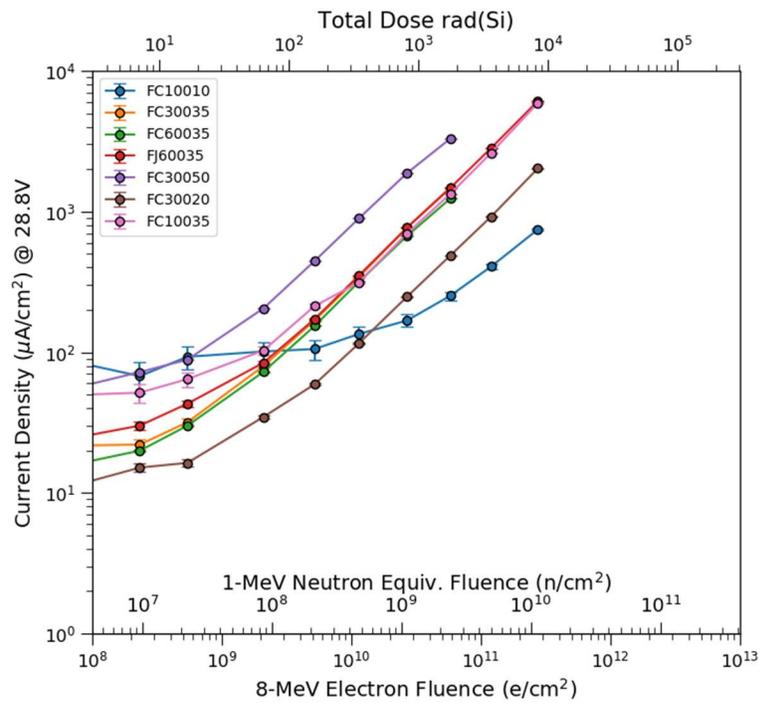

Figure B9. Current Density as a function of dose\fluence at 28.8V bias for alls SiPMs.

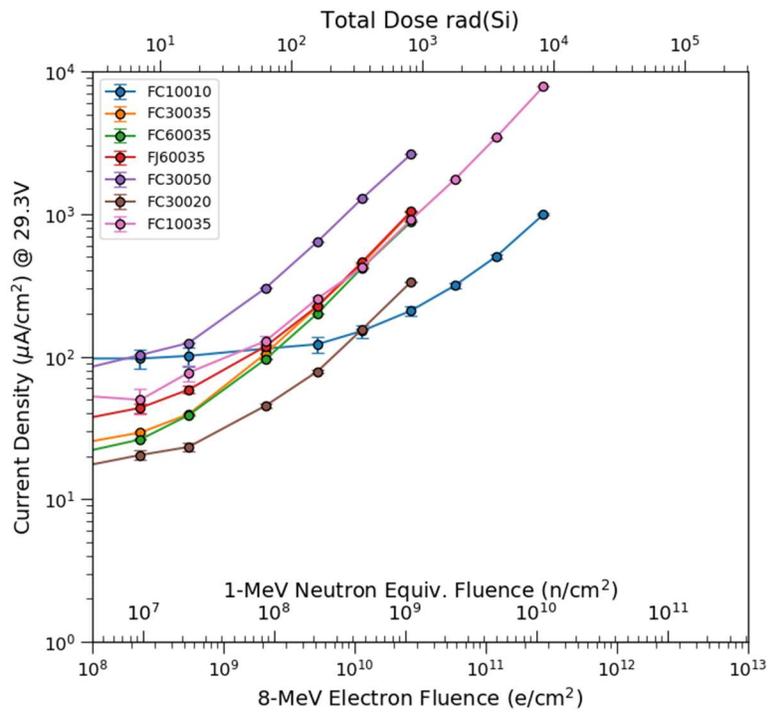

Figure B10. Current Density as a function of dose\fluence at 29.3V bias for alls SiPMs.

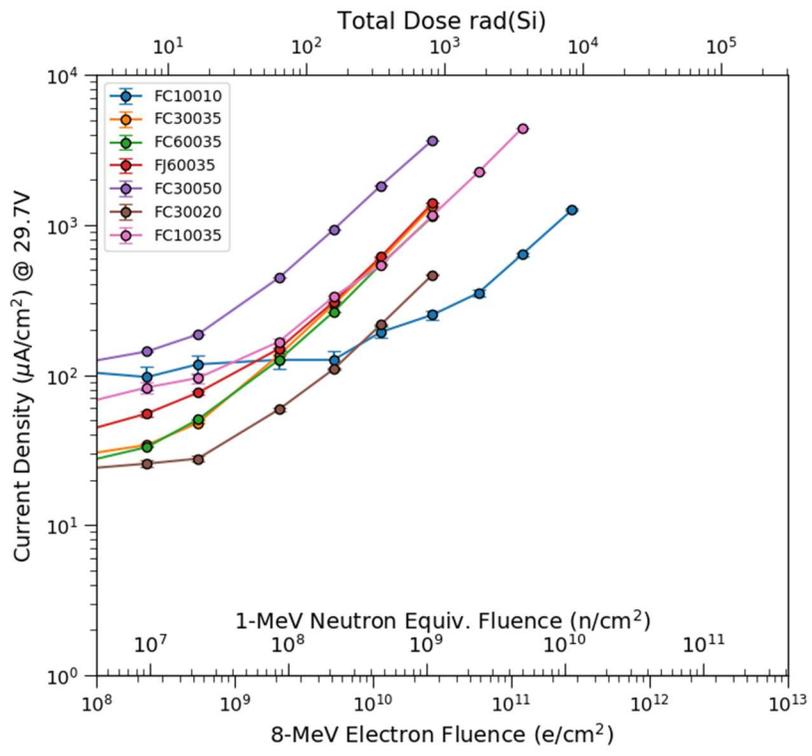

Figure B11. Current Density as a function of dose\fluence at 29.7V bias for alls SiPMs.

**Appendix C. I-V Curves for Proton Irradiated Samples.**

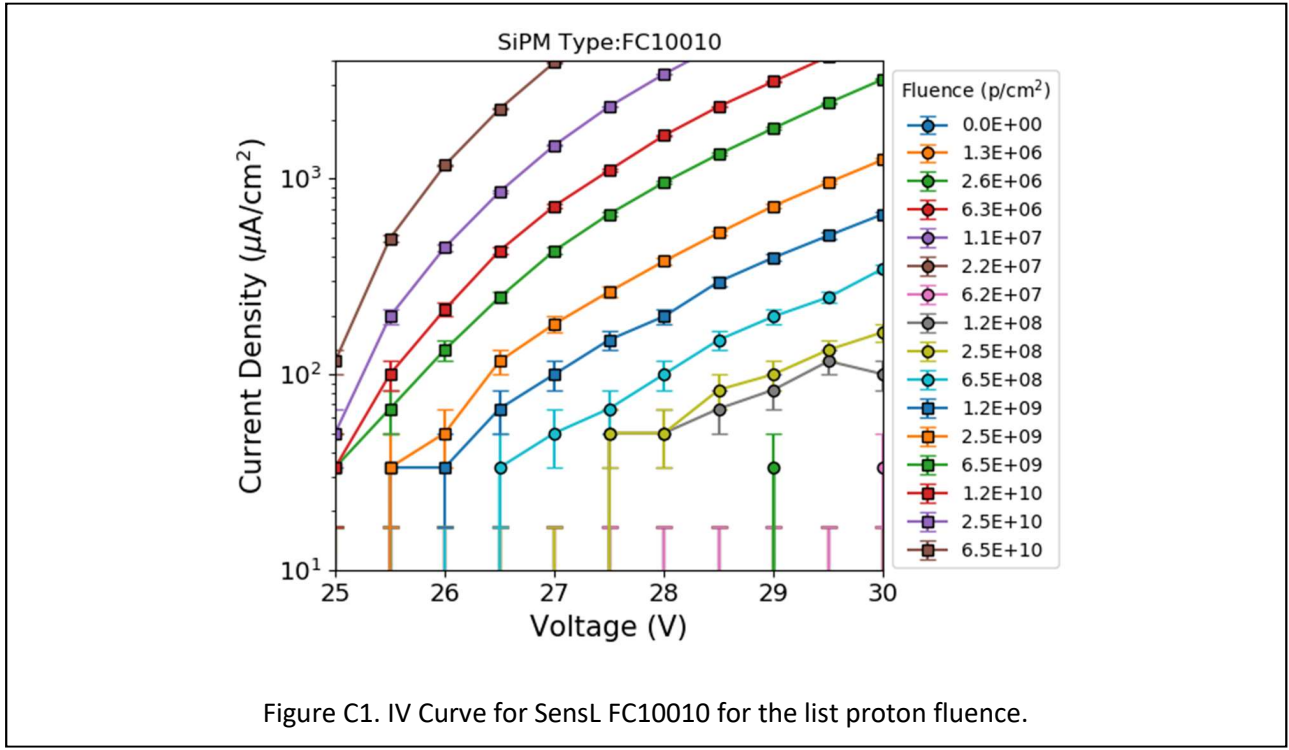

Figure C1. IV Curve for SensL FC10010 for the list proton fluence.

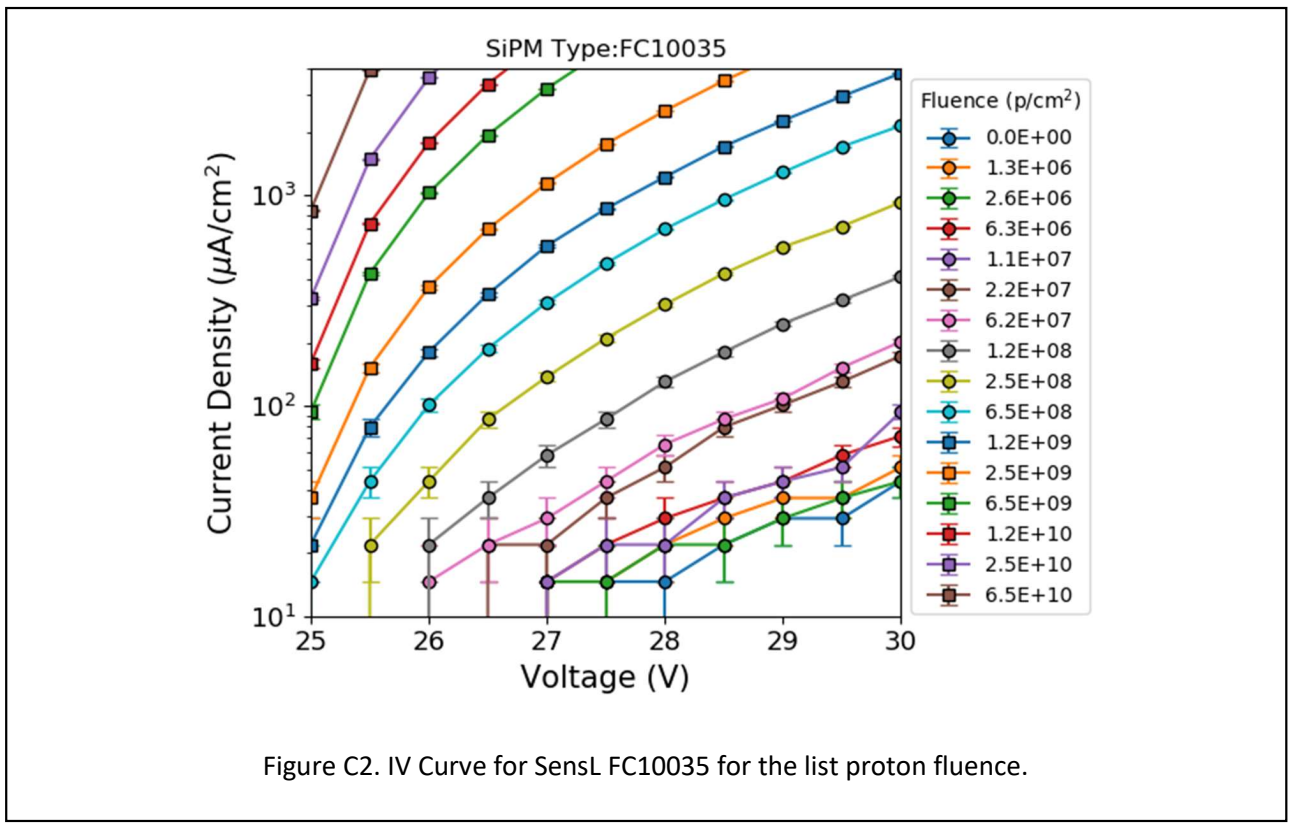

Figure C2. IV Curve for SensL FC10035 for the list proton fluence.

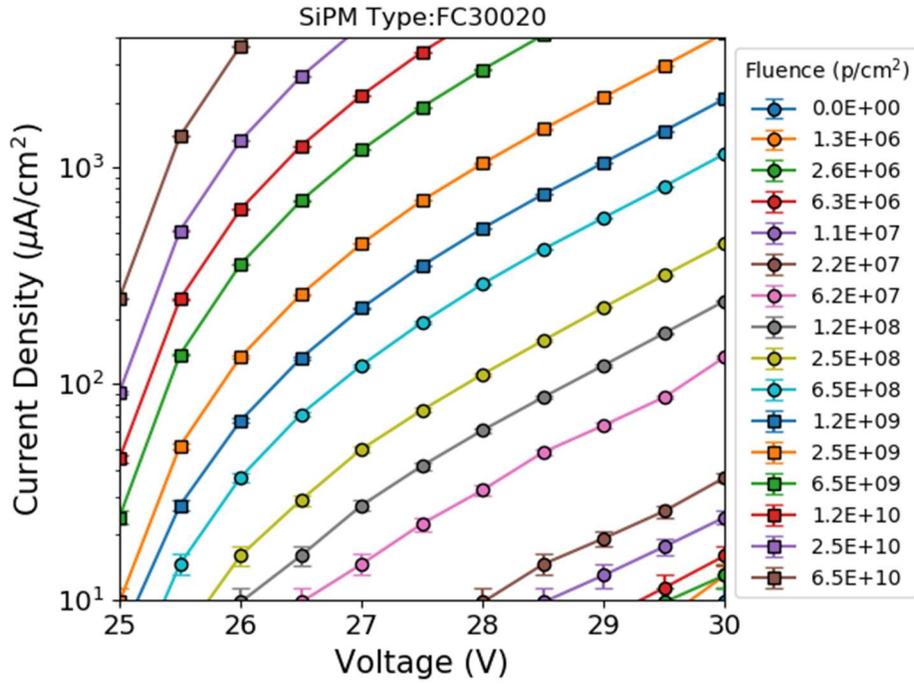

Figure C3. IV Curve for SensL FC30020 for the list proton fluence.

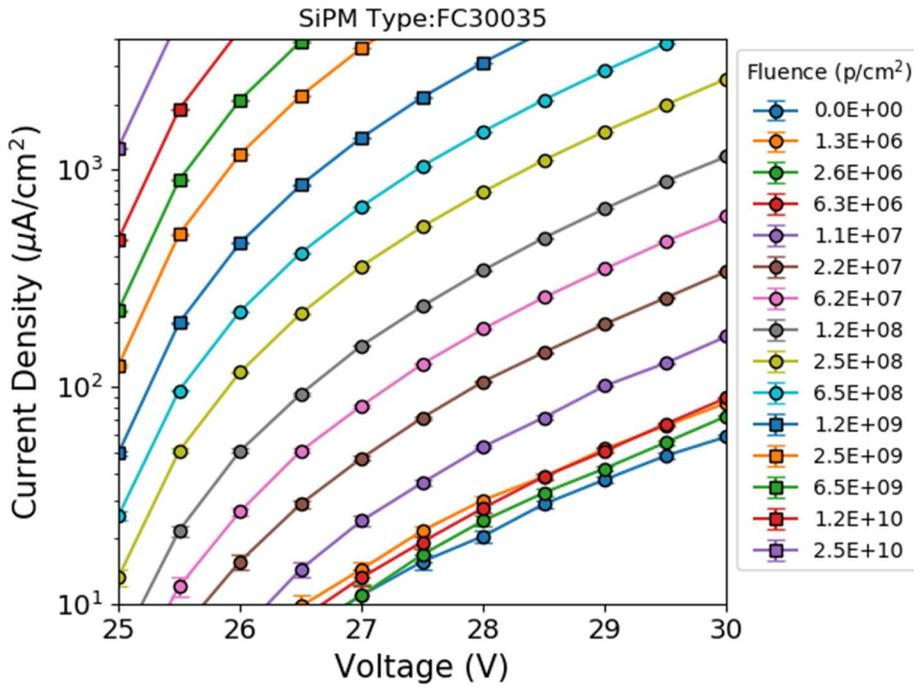

Figure C4. IV Curve for SensL FC30035 for the list proton fluence.

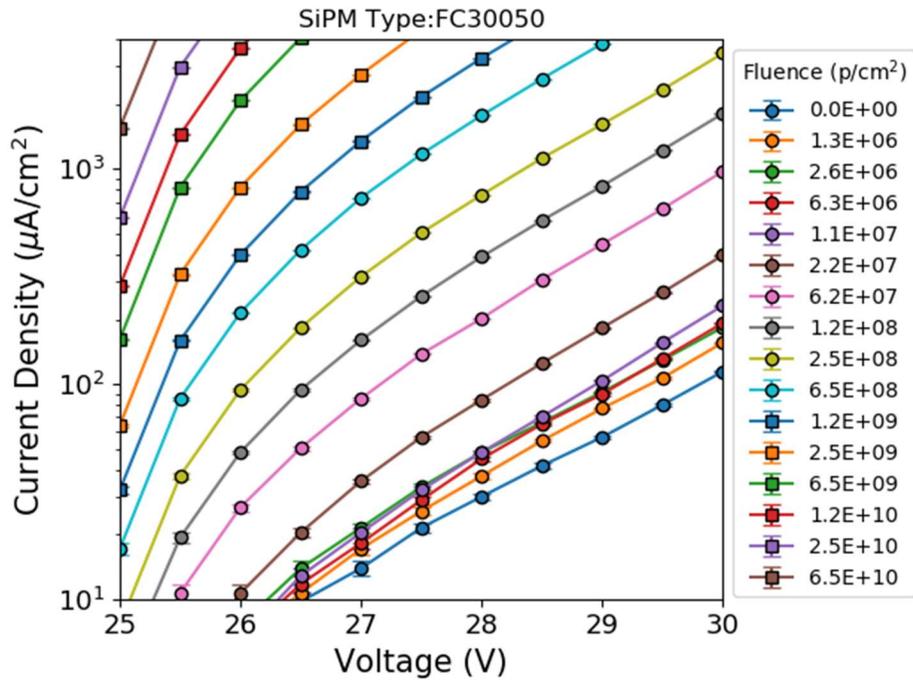

Figure C5. IV Curve for SensL FC30050 for the list proton fluence.

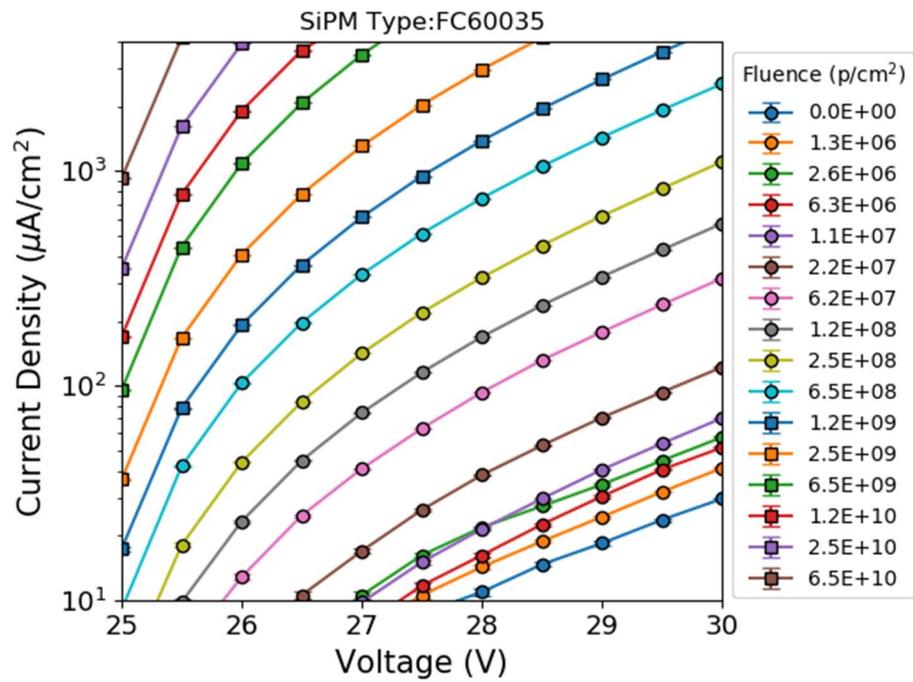

Figure C6. IV Curve for SensL FC60035 for the list proton fluence.

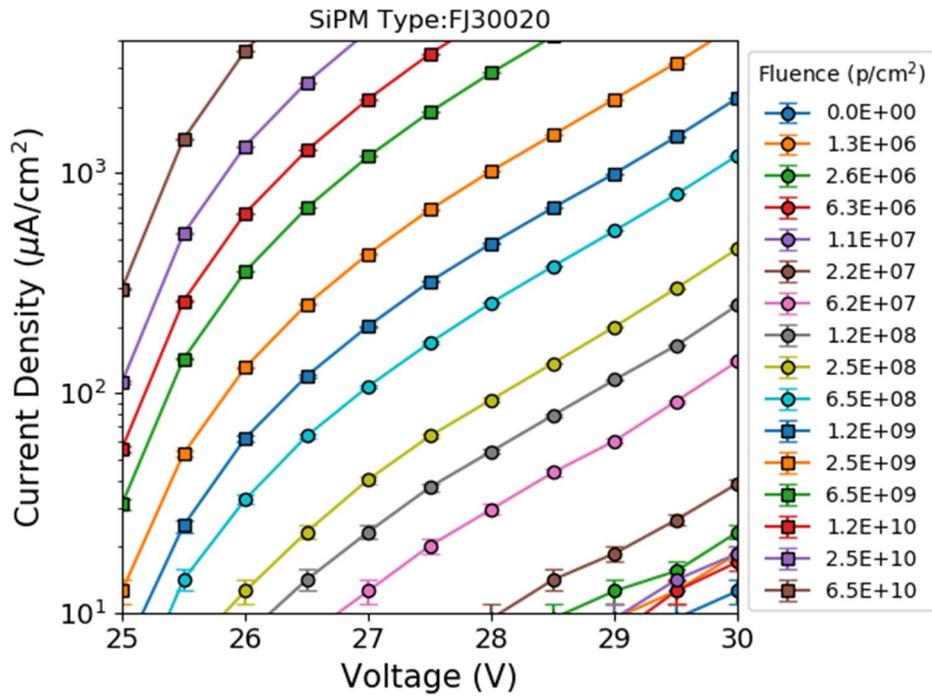

Figure C7. IV Curve for SensL FJ30020 for the list proton fluence.

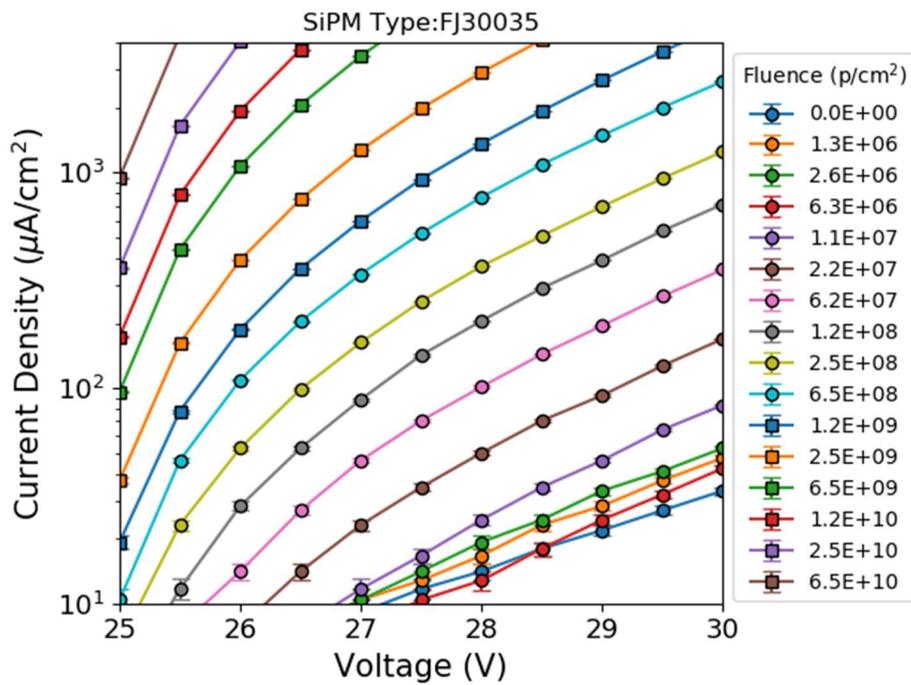

Figure C8. IV Curve for SensL FJ30035 for the list proton fluence.

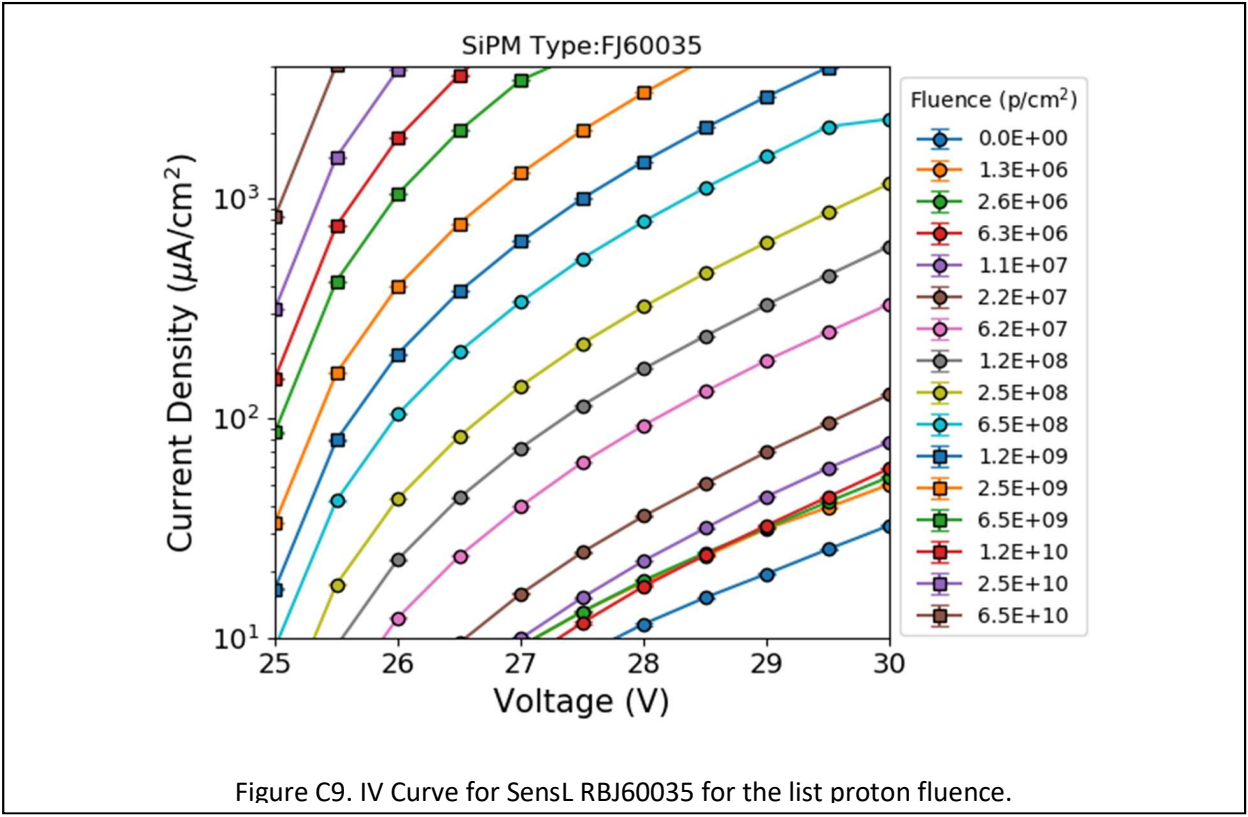

Figure C9. IV Curve for SensL RBJ60035 for the list proton fluence.

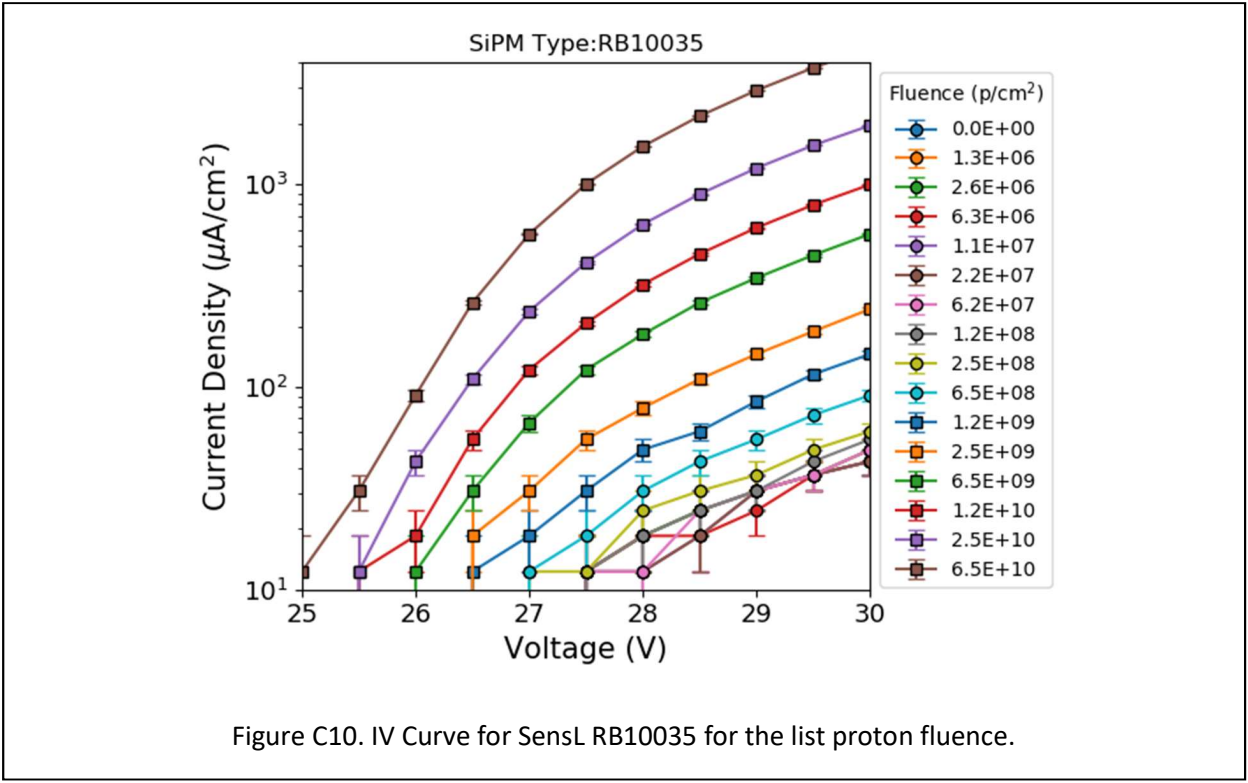

Figure C10. IV Curve for SensL RB10035 for the list proton fluence.

# Appendix D. Current Density as a Function of Dose for Proton Irradiated Samples

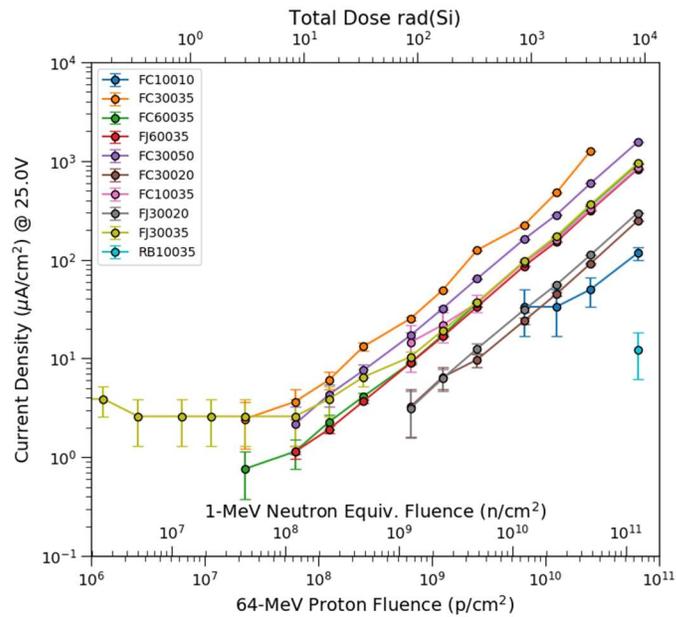

Figure D1. Current Density as a function of dose\fluence at 25V bias for alls SiPMs.

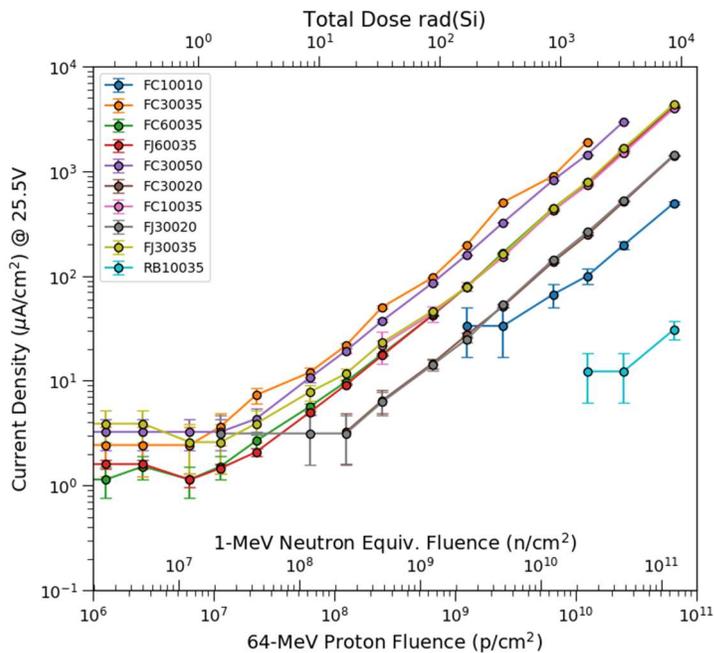

Figure D2. Current Density as a function of dose\ fluence at 25.5V bias for alls SiPMs.

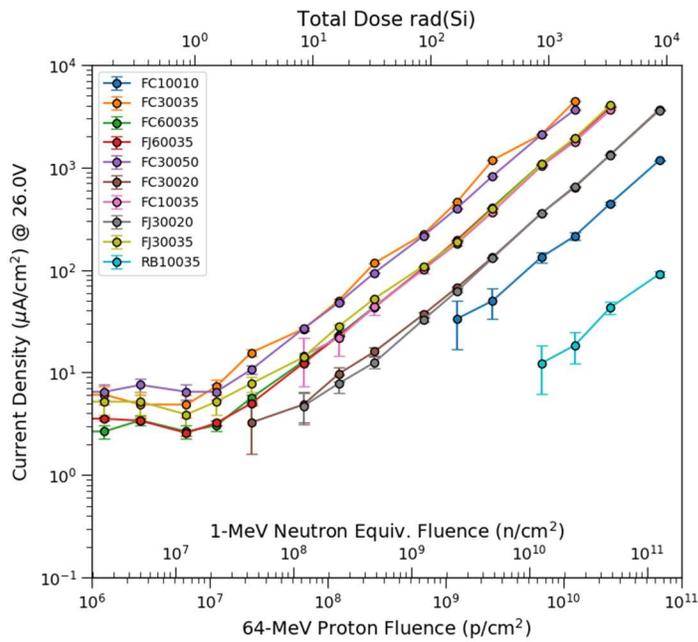

Figure D3. Current Density as a function of dose\ fluence at 26.0V bias for alls SiPMs.

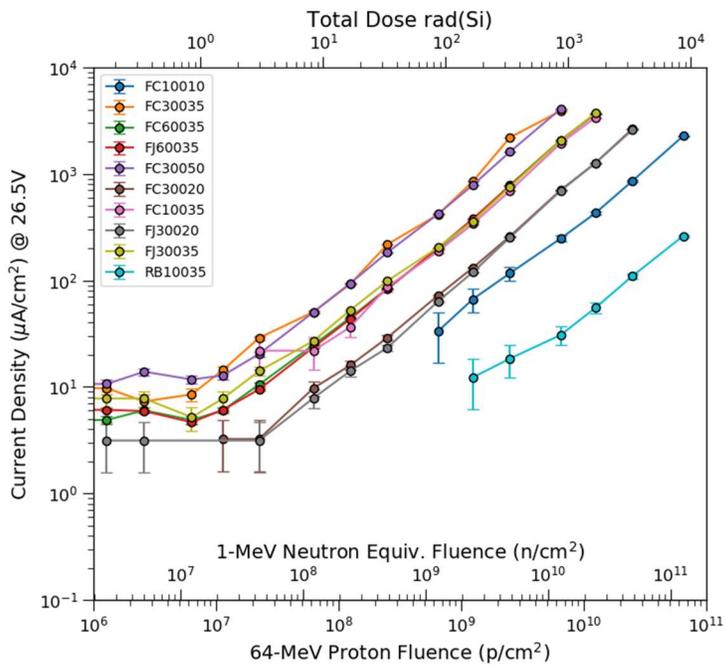

Figure D4. Current Density as a function of dose\ fluence at 26.5V bias for alls SiPMs.

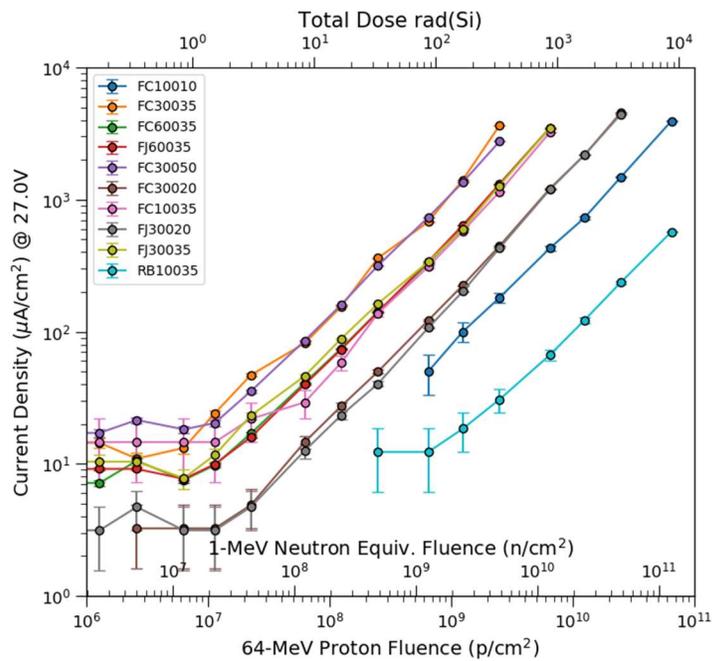

Figure D5. Current Density as a function of dose\ fluence at 27.0V bias for alls SiPMs.

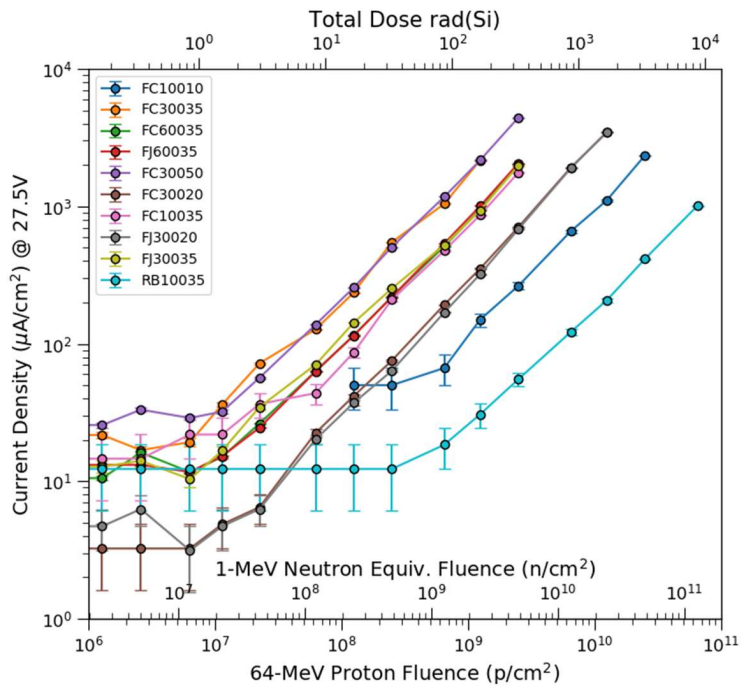

Figure D6. Current Density as a function of dose\ fluence at 27.5V bias for alls SiPMs.

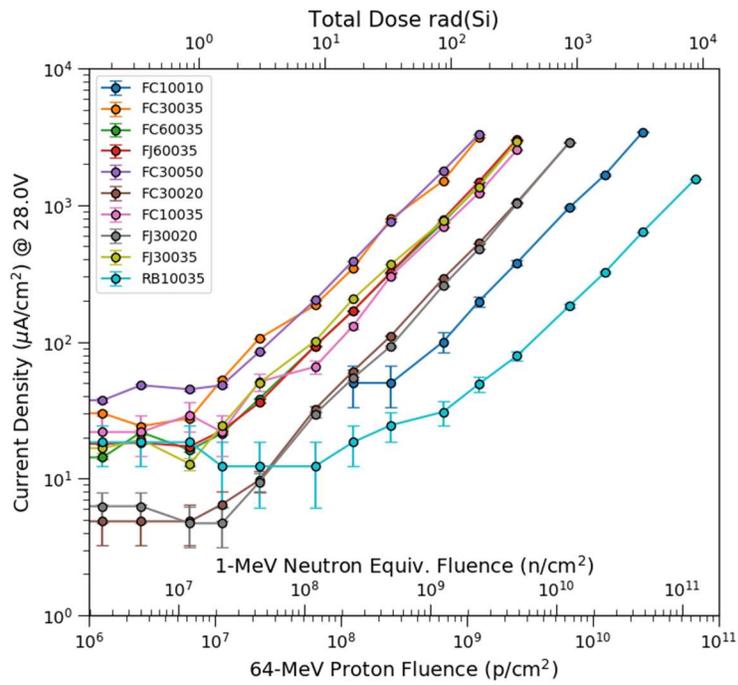

Figure D7. Current Density as a function of dose\ fluence at 28.0V bias for alls SiPMs.

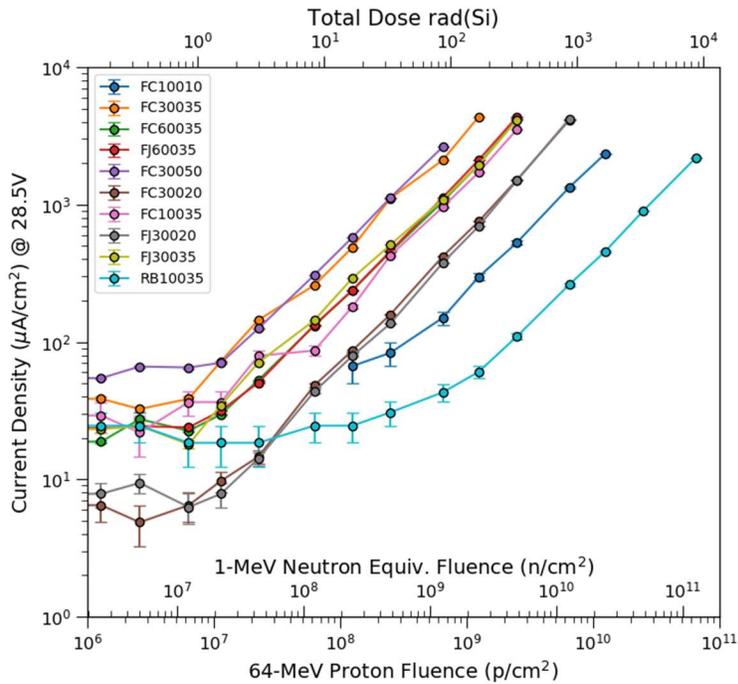

Figure D8. Current Density as a function of dose\ fluence at 28.5V bias for alls SiPMs.

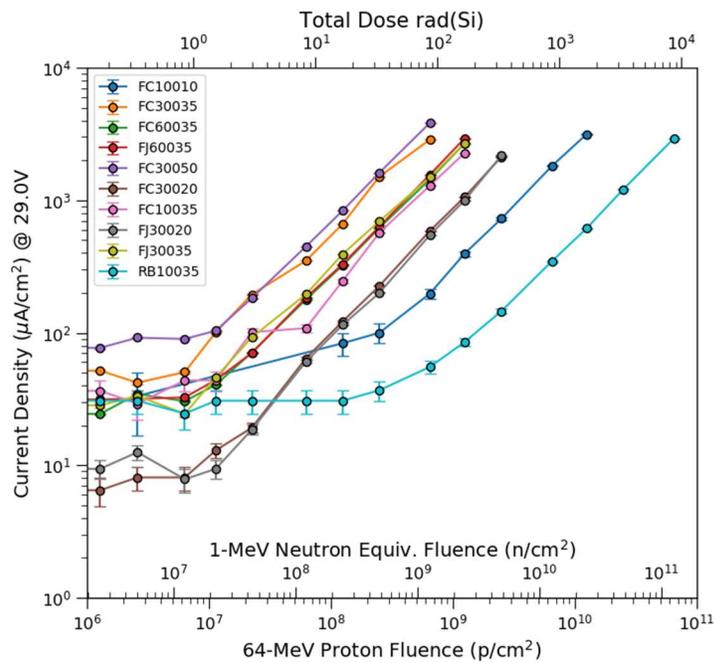

Figure D9. Current Density as a function of dose\ fluence at 29.0V bias for alls SiPMs.

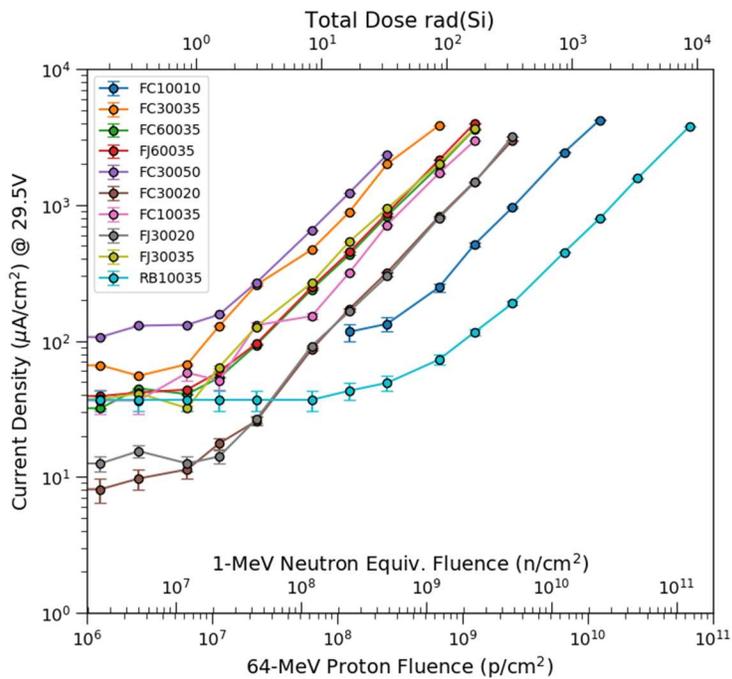

Figure D10. Current Density as a function of dose\ fluence at 29.5V bias for alls SiPMs.

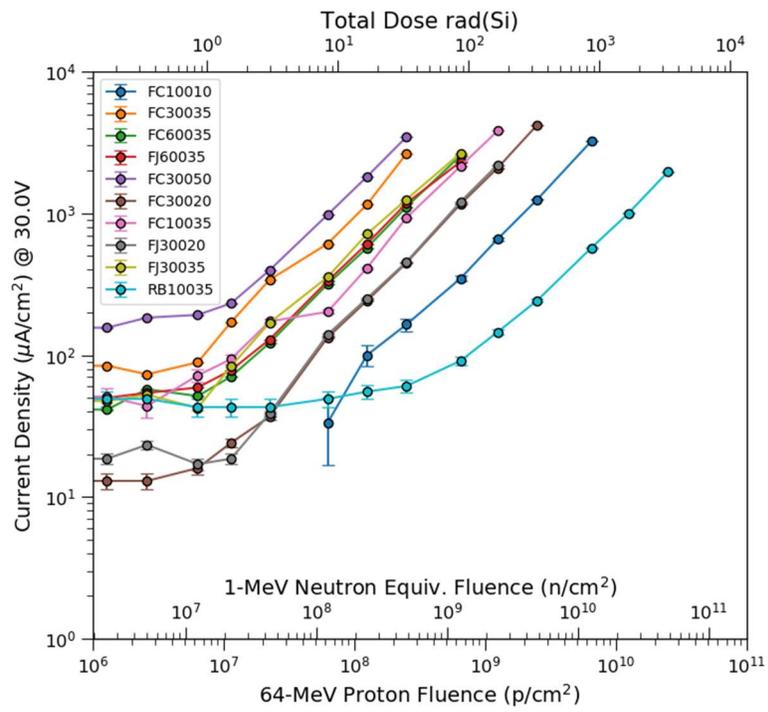

Figure D11. Current Density as a function of dose\ fluence at 30.0V bias for alls SiPMs.